\documentclass[aps,PRD,twocolumn,nofootinbib,11,unsortedaddress,superscriptaddress]{revtex4-1}

\usepackage{amssymb} 
\usepackage{amsmath} 
\usepackage{times}

\usepackage{txfonts}

\usepackage[pdftex]{graphicx}
\usepackage{xspace}

\newcommand{\jpsi}{\ensuremath{\mathrm{J}/\psi}\xspace}

\newcommand{\ups}{\ensuremath{\Upsilon}\xspace}

\newcommand{\HELACOnia}{{\sc\small HELAC-Onia}\xspace}
\newcommand{\Pythia}{{\sc\small Pythia}\xspace}
\newcommand{\MGaMC}{{\sc\small MadGraph5\_aMC@NLO}}
\newcommand{\ie}{{\it i.e.}}

\newcommand{\Q}{{\cal Q}}
\newcommand{\bq}{\begin{equation}}
\newcommand{\eq}{\end{equation}}
\newcommand{\bqa}{\begin{eqnarray}}
\newcommand{\eqa}{\end{eqnarray}}

\newcommand{\snn}{\ensuremath{\sqrt{s_{\mathrm{\scriptscriptstyle NN}}}}\xspace}

\newcommand{\ce}[1]{Eq.~(\ref{#1})}

\newcommand{\cf}[1]{{Fig.~\ref{#1}}}

\newcommand{\ct}[1]{{Table~(\ref{#1})}}

\def\RpPb    {\mbox{$R_{p+\rm Pb}$}}
\def\Ncoll   {\mbox{$N_{\rm coll}$}}

\usepackage{rotating}
\usepackage{lineno}
\modulolinenumbers[5]

\usepackage{color}

 \usepackage{ifpdf}
 \ifpdf
 \usepackage[pdftex]{hyperref}
 \else
 \usepackage[hypertex]{hyperref}
 \fi

 \hypersetup{
   pdftitle={},%
   pdfauthor={},%
   pdfsubject={},%
   pdfkeywords={},%
   pdfstartview={},%
   bookmarksopen=true, breaklinks=true, debug=true, %
   colorlinks=true, linkcolor=red, citecolor=blue, urlcolor=blue
 }

\begin{document}%
%
%
%
%
\title{Feasibility studies for quarkonium production at a fixed-target experiment using the LHC proton and lead beams (AFTER@LHC)}

\date{\today}

\author{L. Massacrier}
\affiliation{LAL, Universit\'e Paris-Sud, CNRS/IN2P3, F-91406, Orsay, France}
\affiliation{IPNO, Universit\'e Paris-Sud, CNRS/IN2P3, F-91406, Orsay, France}

\author{B. Trzeciak}
\affiliation{FNSPE, Czech Technical U., Prague, Czech Republic}

\author{F.~Fleuret}
\affiliation{Laboratoire Leprince Ringuet, \'Ecole Polytechnique, CNRS/IN2P3,  91128 Palaiseau, France}

\author{C.~Hadjidakis}
\affiliation{IPNO, Universit\'e Paris-Sud, CNRS/IN2P3, F-91406, Orsay, France}

\author{D.~Kikola}
\affiliation{Faculty of Physics, Warsaw University of Technology, ul. Koszykowa 75, 00-662 Warsaw, Poland}

\author{J.P.~Lansberg}
\affiliation{IPNO, Universit\'e Paris-Sud, CNRS/IN2P3, F-91406, Orsay, France}

\author{H.-S. Shao}
\affiliation{PH Department, TH Unit, CERN, CH-1211, Geneva 23, Switzerland}

\date{\today}

\begin{abstract}
Used in the fixed-target mode, the multi-TeV LHC proton and lead beams allow for studies of heavy-flavour hadroproduction
with unprecedented precision at backward rapidities --far negative Feyman-$x$-- using conventional detection techniques.  At the nominal LHC energies, 
quarkonia can be studied in detail in $p+p$, $p+d$ and $p+A$ collisions at $\sqrt{s_{NN}} \simeq 115$~GeV as well as in
${\rm Pb}+p$ and ${\rm Pb}+A$ collisions at $\sqrt{s_{NN}} \simeq 72$~GeV with luminosities roughly equivalent to that of the collider mode,
\ie~ up to 20~fb$^{-1}$yr$^{-1}$ in $p+p$ and $p+d$ collisions, up to 0.6 fb$^{-1}$yr$^{-1}$ in $p+A$ collisions and 
up to 10 nb$^{-1}$yr$^{-1}$ in ${\rm Pb}+A$ collisions. 
In this paper, we assess the feasibility of such studies by performing fast simulations using the performance of a 
LHCb-like detector. 
\end{abstract}

\maketitle

\section{Introduction}
\label{sec:intro}

Since its start-up, the large hadron collider (LHC) --the most energetic 
hadron collider ever built so far-- has already made the demonstration of its outstanding
capabilities. These can greatly be complemented by the addition of a fixed-target physics
program. Its multi-TeV beams indeed allow one to study $p+p$, $p+d$ and $p+A$ collisions at 
a center-of-mass energy $\sqrt{s_{NN}} \simeq 115$~GeV as well as ${\rm Pb}+p$ and ${\rm Pb}+A$ 
collisions at $\sqrt{s_{NN}} \simeq 72$~GeV, with the high precision typical of the 
fixed-target mode. In this context the proposal  of a fixed 
target experiment at the LHC~\cite{Brodsky:2012vg}, referred to
AFTER@LHC -- A Fixed Target Experiment --, has been promoted~\cite{Brodsky:2012vg} in order
to complement the existing collider experiments such as the Relativistic Heavy Ion collider (RHIC) 
or the future Electron-Ion Collider (EIC) project in a similar energy range. The idea underlying the AFTER@LHC proposal 
is a multi-purpose detector allowing  for the study of a multitude of probes. 

Various technological ways to perform fixed-target 
experiment at the LHC exist. On the one hand, the beam can be extracted by means of a bent crystal. 
This technology ~\cite{Arduini:1997kh,Scandale:2011za} is currently developed as a smart beam-collimation 
solution and is studied by the UA9/LUA9 collaboration respectively at SPS and LHC. 
A bent crystal installed in the halo of the LHC beam would deflect the particles of the halo 
onto a target, with a flux of 5 $\times 10^{8}$ proton/s without any impact on the LHC 
performances~\cite{Uggerhoj:2005xz,Scandale:2011za,LHCC2011}. 

On the other hand, the LHC beam 
can go through an internal-gas-target system in an existing (or new) LHC experiment. Such 
a system is already tested at low gas pressure by the LHCb collaboration to monitor the 
luminosity of the beam \cite{Barschel:2014iua,FerroLuzzi:2005em,Aaij:2014ida}. Data were 
taken at a center-of-mass energy of  $\sqrt{s_{NN}}$ = 87 (54) GeV with $p+{\rm Ne}$ (${\rm Pb}+{\rm Ne}$) collisions 
during pilot runs in 2012 and 2013. Although this system, called SMOG, was tested during only 
few hours in a row, no decrease of the LHC performances was observed. 

In the bent-crystal case, the luminosity achievable with AFTER@LHC would surpass that of RHIC by 
3 order of magnitudes~\cite{Brodsky:2012vg}. We have reported in \ct{tablumi} the instantaneous and yearly 
integrated luminosities expected with the proton and Pb beams on various target species of 
various thicknesses, for the bent-crystal as well as internal-gas-target options. Integrated 
luminosities as large as 20 fb$^{-1}$ can be delivered during a one-year run of $p+{\rm H}$ collisions with 
a bent crystal. Besides, it is worth mentioning that both technologies allow one to polarise the 
target, which is an important requirement to lead an extensive spin physics 
programme~\cite{Brodsky:2012vg,Rakotozafindrabe:2013au}. 

Overall, thanks to the large luminosity expected, AFTER@LHC 
would become a quarkonium~\cite{Brambilla:2010cs}, prompt photon and heavy-flavour 
observatory~\cite{Brodsky:2012vg,Lansberg:2012kf} in $p+p$ and $p+A$ collisions 
where, by instrumenting the target-rapidity region, gluon and heavy-quark distributions of the proton, 
the neutron and the nuclei can be accessed at large $x$ and even at $x$ larger than unity in 
the nuclear case~\cite{Lansberg:2013wpx}. In addition, the fixed-target mode allows for 
single-target-spin-asymmetry measurements over the full backward 
rapidity domain up to $x_F \simeq - 1$~\cite{Liu:2012vn,Kanazawa:2015fia}. 
Also, the versatility in the target choices offer a unique opportunity to study the 
nuclear matter versus the hot and dense matter formed in heavy-ion collisions which can be studied during 
the one-month lead run. In the latter case, modern detection technology (such as high granularity calorimeter) 
should allow for extensive studies of quarkonium excited states, from the $\psi(2S)$ to the $\chi_c$ and $\chi_b$ resonances 
thanks to the boost of the fixed-target mode~\cite{Rakotozafindrabe:2012ei}.



\begin{table}
\begin{center}
\caption{\label{tablumi}Expected luminosities obtained for a 7 (2.76) TeV proton (Pb) beam extracted by means of a bent crystal or obtained with an internal gas target.}
{\begin{tabular}{|c c c c c c|}
  \hline
  Beam & Target & Thickness & $\rho$ & $\cal{L}$ & $\int{\cal{L}}$  \\ 
   &  & (cm) & (g.cm$^{-3}$) & ($\mu$b$^{-1}$.s$^{-1}$) & (pb$^{-1}$.y$^{-1})$  \\ \hline
  p & Liquid H & 100 & 0.068 & 2000 & 20000 \\
  p & Liquid D & 100 & 0.16 & 2400 & 24000 \\
  p & Pb & 1 & 11.35 & 16 & 160 \\ \hline 
  Pb & Liquid H & 100 & 0.068 & 0.8 & 0.8 \\
  Pb & Liquid D & 100 & 0.16 & 1 & 1 \\
  Pb & Pb & 1 & 11.35 & 0.007 & 0.007 \\ \hline \hline
  Beam & Target & Usable gas zone & Pressure &  $\cal{L}$ & $\int{\cal{L}}$    \\
       &  & (cm) & (Bar) & ($\mu$b$^{-1}$.s$^{-1}$) & (pb$^{-1}$.y$^{-1})$  \\ \hline
    p  & perfect gas & 100 & $10^{-9}$ & 10 & 100 \\ \hline
    Pb & perfect gas & 100 & $10^{-9}$ & 0.001 & 0.001 \\ 
       
  \hline
\end{tabular}}
\end{center}
\end{table} 
In this paper, we report on a feasibility study of quarkonium production 
at a fixed-target experiment using LHC beams. In section II, we outline the simulation framework 
which was used. In section III, we describe how a fast simulation of a detector response 
has been implemented, following a LHCb-like detector setup. In section IV, we present the 
charmonium and bottomonium family studies performed with the $p+{\rm H}$ simulations at $\sqrt{s}$ = 115 GeV.
In section V, we present multiplicity studies in $p+A$ and $A+p$ collisions as well as the 
expected nuclear modification factors for J/$\psi$ and $\Upsilon$ in $p+\mathrm{Pb}$ collisions at 
$\sqrt{s_{NN}}$ = 115 GeV. Finally in section VI some prospects for $\mathrm{Pb}+A$ measurement at $\sqrt{s_{NN}}$ = 72 GeV are given. Section VII 
gathers our conclusions.

\section{Simulation inputs} 
\label{sec:input}

In order to get the most realistic minimum bias simulations at AFTER@LHC energy for quarkonium studies in the 
dimuon decay channels, we have simulated the quarkonium signal and all the background sources separately to have 
under control the transverse momentum and rapidity input distributions as well as  the normalisation of the different sources.

The simulation has been performed for $p+p$ collisions at $\sqrt{s}$~=~115~GeV. 
On the one hand, the quarkonium signal and the correlated background (Drell-Yan, $c\bar{c}$, $b\bar{b}$) were simulated with \HELACOnia~\cite{Shao:2012iz} which produces outputs following the format of Les Houches Event Files~\cite{Alwall:2006yp}. The outputs were then processed with \Pythia (\Pythia~8.185~\cite{Sjostrand:2007gs}) to perform the hadronisation, the initial/final-state radiations and the decay of the resonances. On the other hand, the uncorrelated background was obtained from minimum bias $p+p$ collisions generated with \Pythia. 

The relative normalisation of the signal and background sources was performed according to the 
production cross section of the process (taking into account initial phase space cuts, if any). 
Values of the cross section and the number of simulated events $N_{sim}$ --not to be confused with the expected events for
a specific luminosity-- are reported in 
\ct{tab:crosssection}. The cross section values are integrated over rapidity and $p_{\rm T}$.

\begin{table*}[!hbtp] 
\begin{tabular}{c|p{3.cm}p{3.5cm}}\footnotesize
         & $\sigma_{tot}$ (mb) & $N_{sim}$  \\ \hline \hline
J/$\psi$ & 1.30 $\times$ 10$^{-3}$ & 1.47 $\times$ 10$^{6}$\\ 
$\psi$(2S) & 1.61 $\times$ 10$^{-4}$ & 1.12 $\times$ 10$^{6}$\\
$\Upsilon$ (1S) & 4.30 $\times$ 10$^{-7}$ & 1.46 $\times$ 10$^{6}$ \\
$\Upsilon$ (2S) & 1.22 $\times$ 10$^{-7}$ & 1.49 $\times$ 10$^{6}$ \\
$\Upsilon$ (3S) & 5.28 $\times$ 10$^{-8}$ & 1.48 $\times$ 10$^{6}$ \\
Drell-Yan (M $>$ 2.5 GeV/c$^{2}$) & 2.52 $\times$ 10$^{-6}$ &  4.3 $\times$ 10$^{5}$\\
Drell-Yan (M $>$ 7 GeV/c$^{2}$) & 1.49 $\times$ 10$^{-7}$ & 2.0 $\times$ 10$^{6}$  \\
$c\bar{c}$ &  2.29 $\times$ 10$^{-1}$ & 81.5 $\times$ 10$^{6}$ \\	
$b\bar{b}$ &  4.86 $\times$ 10$^{-4}$ ($gg \rightarrow b\bar{b}$) \newline  1.49 $\times$ 10$^{-4}$ ($q\bar{q} \rightarrow b\bar{b}$) & 32.3 $\times$ 10$^{6}$ ($gg \rightarrow b\bar{b}$) \newline 85.7 $\times$ 10$^{6}$ ($q\bar{q} \rightarrow b\bar{b}$) \\
minimum bias & 26.68 & 11.0 $\times$ 10$^{8}$\\ 
\end{tabular}
\caption{Total cross section for different processes in $p+p$ collisions at $\sqrt{s}$~=~115~GeV and number of simulated events $N_{sim}$.} 
\label{tab:crosssection}
\end{table*}

\subsection{Signal and correlated background}

\subsubsection{Quarkonium signal}

J/$\psi$, $\psi$(2S), $\Upsilon(1S)$, $\Upsilon(2S)$ and $\Upsilon(3S)$ were simulated in a data-driven way. The amplitude of $gg \rightarrow \Q+X$  (where $\Q$ is the quarkonium) is expressed in an empirical functional form~\cite{Kom:2011bd}:
\bqa
&&\overline{|\mathcal{A}_{gg\rightarrow\Q+X}|^2}=\nonumber\\
&&\left\{
\begin{array}{ll}
K\exp(-\kappa\frac{p_\mathrm{T}^2}{M_{\Q}^2})
& \mbox{when $p_\mathrm{T}\leq \langle p_\mathrm{T}\rangle$} \\
K\exp(-\kappa\frac{\langle p_\mathrm{T} \rangle^2}{M_{\Q}^2})\left(1+\frac{\kappa}{n}\frac{p_\mathrm{T}^2-\langle p_\mathrm{T} \rangle^2}{M_{\Q}^2}\right)^{-n}
& \mbox{when $p_\mathrm{T}> \langle p_\mathrm{T}\rangle$} \\
\end{array}
\right.\label{eq:crystalball}
\eqa
where $K=\lambda^2\kappa\hat{s}/M_{\Q}^2$ with $\hat{s}$ the partonic center-of-mass energy and $M_{\Q}$ the mass of the quarkonium $\Q$ taken from the PDG table~\cite{Agashe:2014kda}.

\begin{figure}[!htb]
		\centering
		\includegraphics[width=0.95\columnwidth]{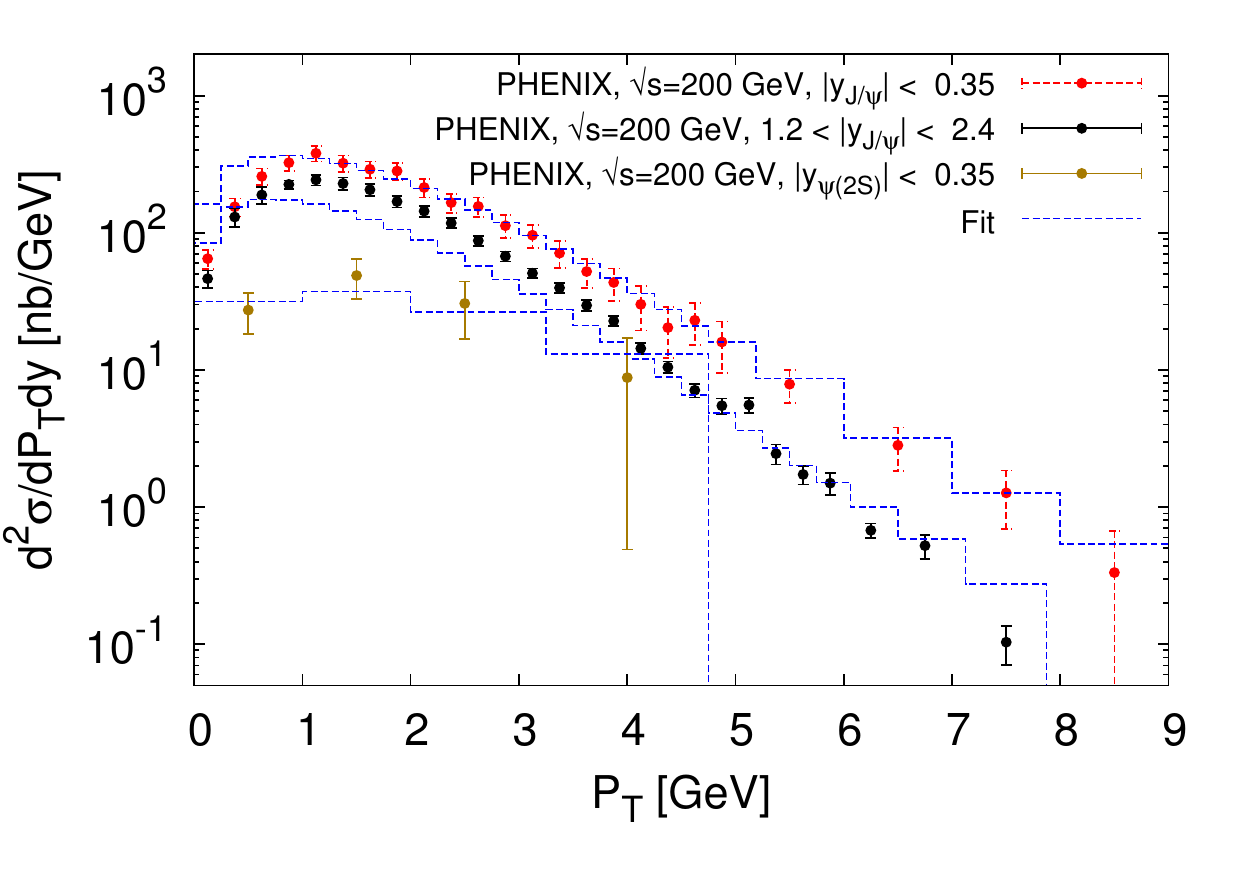}\\
                \includegraphics[width=0.95\columnwidth]{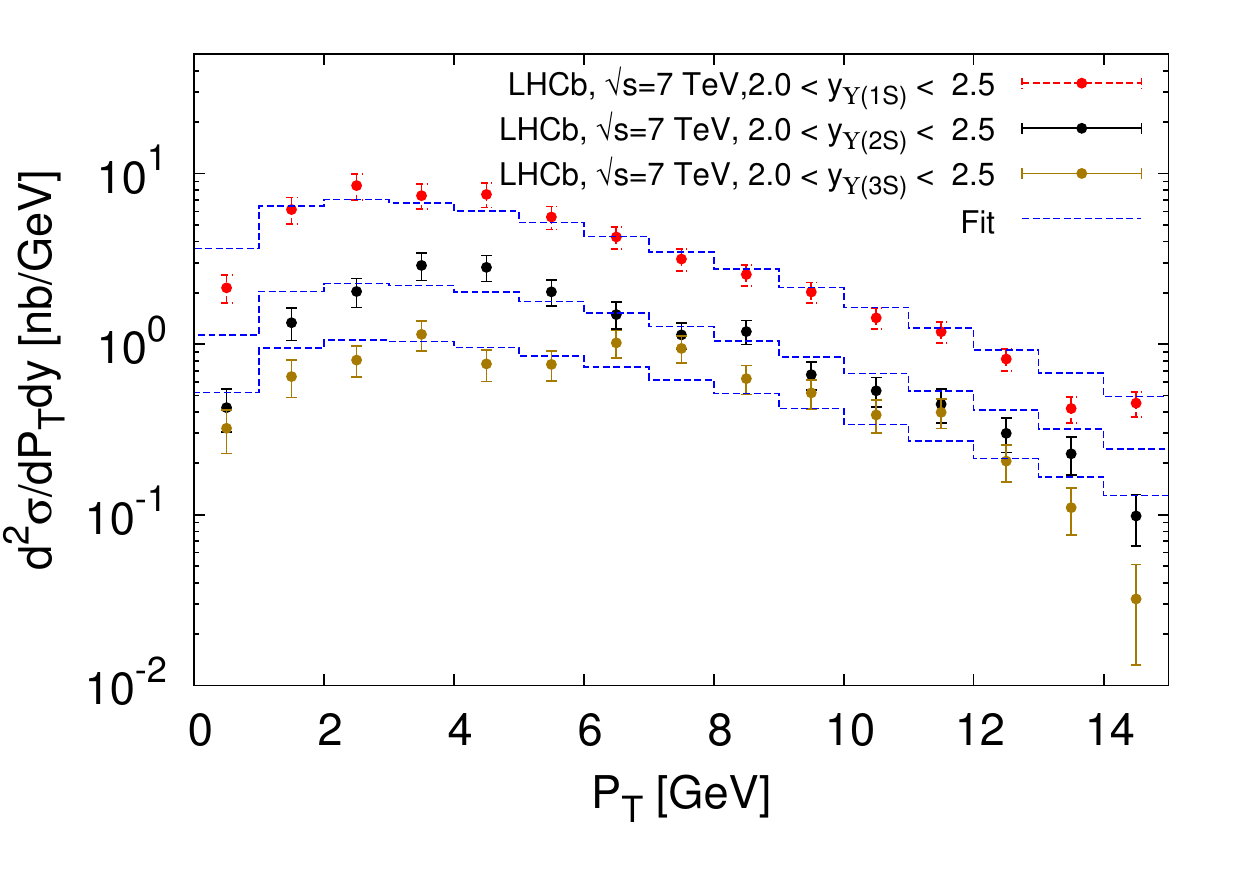}
		\caption{Some illustrative comparison between fits and the PHENIX data~\cite{Adare:2011vq} for charmonium production (upper panel) and LHCb data~\cite{LHCb:2012aa} for bottomonium production (bottom panel).}
		\label{fig:CompareData}
\end{figure}

The parameters $\kappa$, $\lambda$, $n$ and $\langle p_\mathrm{T} \rangle$ were determined by fitting the differential cross section $d^2\sigma/dp_\mathrm{T}dy$ to the experimental data. The dedicated codes used to perform the fit and to generate unweighted events for quarkonium production have been implemented in \HELACOnia~\cite{Shao:2012iz} and we used MSTW2008NLO PDF set~\cite{Martin:2009iq} provided in LHAPDF5~\cite{Whalley:2005nh} and the factorisation scale $\mu_F=\sqrt{M_{\Q}^2+p_\mathrm{T}^2}$. In order to constrain the non-trivial energy dependence of quarkonium production, we used the differential measurements of charmonium production performed by the PHENIX collaboration at RHIC, in $p+p$ collisions at $\sqrt{s}=200$~GeV~\cite{Adare:2011vq} to predict the corresponding yields at $\sqrt{s} = 115$~GeV. Given the lack of such measurements for $\Upsilon$ at RHIC, we performed a combined fit to CDF~\cite{Acosta:2001gv}, ATLAS~\cite{Aad:2012dlq}, CMS~\cite{Chatrchyan:2013yna} and LHCb~\cite{LHCb:2012aa,Aaij:2013yaa} data on $\Upsilon$ production. The values of the fitted parameters are listed in \ct{cbfit}. For illustration, the comparison between fits and the selected experimental data is shown in \cf{fig:CompareData}. 

\begin{table}[!hbtp]
\begin{center}
\begin{tabular}{c|cccc}\footnotesize
          & $\kappa$ &$\lambda$ & \# of data points & $\chi^2$ \\
\hline\hline
$J/\psi$ & $0.674$  & $0.380$ & $51$ & $422$\\
$\psi(2S)$  &  $0.154$ & $0.351$ & $4$ & $1.12$ \\
$\Upsilon(1S)$ & $0.707$  & $0.0837$ & $288$ & $1883$\\
$\Upsilon(2S)$  &  $0.604$ & $0.0563$ & $205$ & $856$ \\
$\Upsilon(3S)$  &  $0.591$ & $0.0411$ & $197$ & $886$ \\
\end{tabular}
\end{center}
\caption{Fit parameters obtained after a combined fit of $d^2\sigma/dp_\mathrm{T}dy$ to the PHENIX data~\cite{Adare:2011vq} for charmonium production and to CDF~\cite{Acosta:2001gv}, ATLAS~\cite{Aad:2012dlq}, CMS~\cite{Chatrchyan:2013yna} and LHCb~\cite{LHCb:2012aa,Aaij:2013yaa} data for bottomonium production. We have fixed $n=2$ and $\langle p_\mathrm{T}\rangle =$~4.5~(13.5)~GeV/$c$ for charmonium (bottomonium) production. The number of fitted data points is also reported.}
\label{cbfit}
\end{table}

In order to increase the statistics of the simulated data sample, the decay of the quarkonium in \Pythia is forced into the dimuon decay channel. The simulated yields are then weighted by the cross section for this process multiplied by the Branching Ratio (BR).  

\subsubsection{Open charm}

Open-charm production was simulated with the process $gg \rightarrow c\bar{c}$ in \HELACOnia. In order to avoid the huge theoretical uncertainties in the state-of-the-art perturbative calculations, open charm yields at $\sqrt{s} = 115$~GeV are also computed in a data-driven way following the method described in the previous section. Similarly, the matrix element of $gg \rightarrow c\bar{c}$ is determined using \ce{eq:crystalball}. The parameters are obtained from a fit to the $p_\mathrm{T}$-differential $c\bar{c}$ cross section measured by the STAR experiment~\cite{Ye:2014eia} in $p+p$ collisions at $\sqrt{s} = 200$~GeV (see \cf{fig:STARccbar}). We obtained $\kappa=0.437$, $\lambda=3.04$ and $\langle p_\mathrm{T} \rangle=2.86$~GeV/$c$ when $n=2$ by using CTEQ6L1~\cite{Pumplin:2002vw} and by fixing the $c$ quark mass to $m_c=1.5$~GeV/$c^2$ and the factorisation scale to $\mu_F=\sqrt{m_c^2+p_\mathrm{T}^2}$. The $\chi^2$ of the fit is equal to 4.39 with 10 experimental data points. The tuned result is shown in \cf{fig:ComparisonSTARccbar}. 
The evolution of the cross section with the energy down to $\sqrt{s} = 115$~GeV is then given by \HELACOnia.

\begin{figure}[!htb]
		\centering
		\includegraphics[width=0.95\columnwidth]{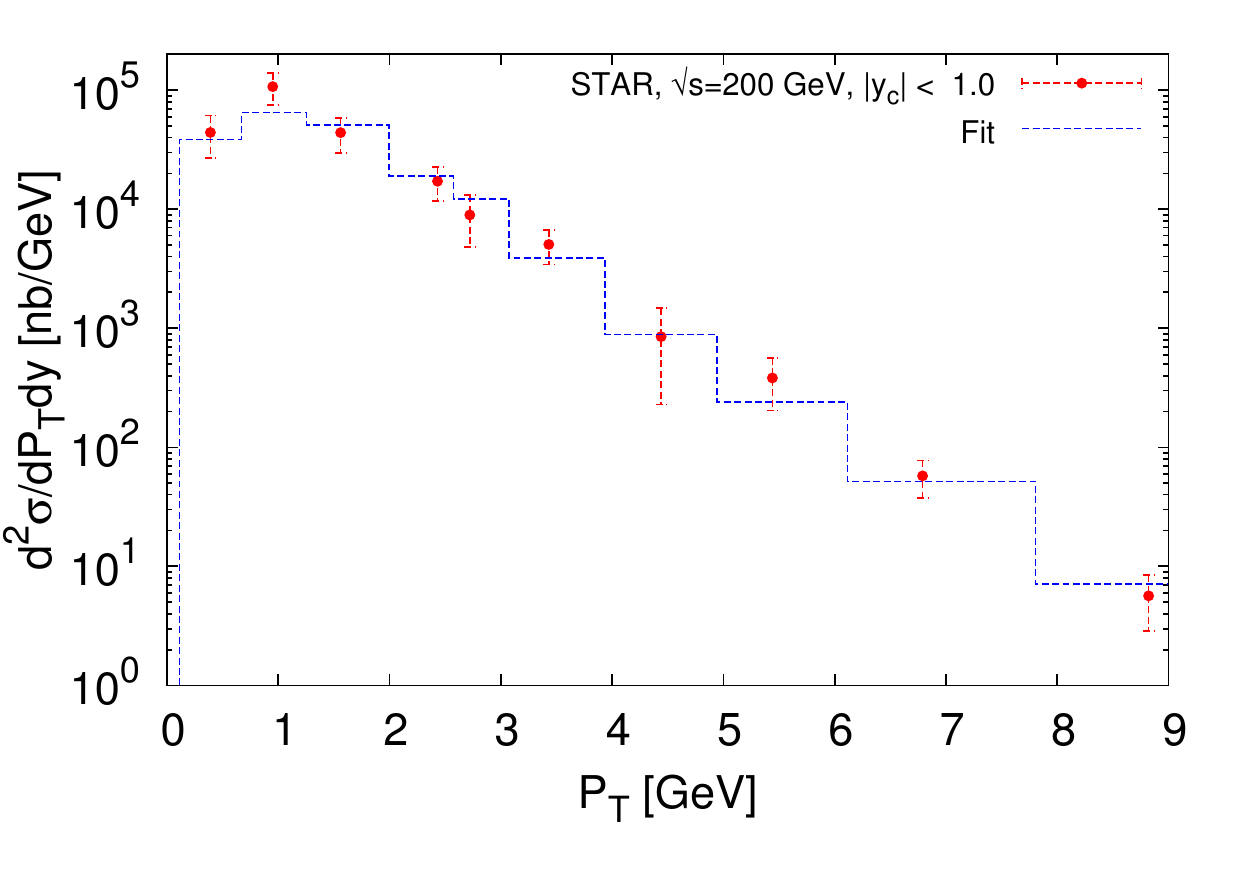}
		\caption{A comparison between fit and the STAR data~\cite{Ye:2014eia} in $p+p$ collisions at $\sqrt{s} = 200$~GeV, for $c\bar{c}$ production.}
		\label{fig:ComparisonSTARccbar}
\end{figure}

After embedding the Les Houches Event File into \Pythia, muons from the underlying \Pythia event can be produced on top of muons from the initial $c\bar{c}$ pair. The combination of those additional muons with a muon from the initial $c\bar{c}$ pair is not included in our definition of open charm correlated background. We have however checked that this contribution is negligible. In order to increase statistics, the $D^{0}$, $\bar{D^{0}}$, $D^{+/-}$ and $D_{s}^{+/-}$ were forced to decay into muons and only those decay muons were considered as correlated background. $\mu^{+}\mu{-}$ pairs coming from all possible combinations: $D^{0}\bar{D^{0}}$, $D^{+}D^{-}$, $D_{s}^{+}D_{s}^{-}$, $D^{0}D^{+/-}$, $D^{0}D_{s}^{+/-}$ and $D^{+/-}D_{s}^{-/+}$ are considered. The simulated events are weighted by the production cross section times the pair Branching Ratio times the fraction of $c$ quark fragmenting to $D^{0},\bar{D^{0}}$, $D^{+/-}$ or $D_{s}^{+/-}$. This fraction is obtained from \Pythia and found to be 95$\%$.

\subsubsection{Open beauty}

The theoretical uncertainty on open beauty production is relatively smaller than the one on open charm production. We therefore calculated 
open-beauty-production yields with a Leading Order (LO) matrix element and which was normalised to the Next-To-Leading-Order (NLO) K factor. 
The NLO cross section with the same setup was calculated by \MGaMC~\cite{Alwall:2014hca}. We used CTEQ6L1 (CTEQ6M) for the LO (NLO) calculation. 
The K factor is found to be 1.83. The renormalisation and factorisation scale is $\mu_R=\mu_F=\sqrt{m_b^2+p_\mathrm{T}^2}$ with the mass of 
the $b$ quark taken as $m_b=4.5$~GeV/$c^2$. We have adopted a similar definition for the open beauty correlated background as the 
one of open charm (see previous section).

\subsubsection{Drell-Yan}

Drell-Yan (DY) correlated background was simulated with the process $q\bar{q} \rightarrow \gamma^{\star}/Z \rightarrow \mu^{+} \mu^{-}$ at LO where $q\bar{q}$ is a pair of same flavour light quarks. The LO calculation was done with the CTEQ6L1 pdf set and the renormalisation and factorisation scale was set to $\mu_R=\mu_F=p_\mathrm{T}$. In order to have enough statistics in the J/$\psi$ and $\psi(2S)$ mass window, a phase space cut requesting that the invariant mass of the dimuons (M) is greater than 2.5~GeV/$c^{2}$ was applied. For the simulation of the DY background under the $\Upsilon$ family peaks, a phase space cut $M > 7 $~GeV/$c^{2}$ was applied. The DY cross section obtained with \HELACOnia at $\sqrt{s} = 38.8$~GeV is compared to the existing E866 data at the same energy~\cite{Webb:2003ps}. A K factor 1.2 is needed to match the data and therefore it was also applied at $\sqrt{s} = 115$~GeV.  Such a K factor is known to approximately account for the higher-order QCD corrections.

\begin{figure}[!htb]
		\centering
		\includegraphics[width=0.9\columnwidth]{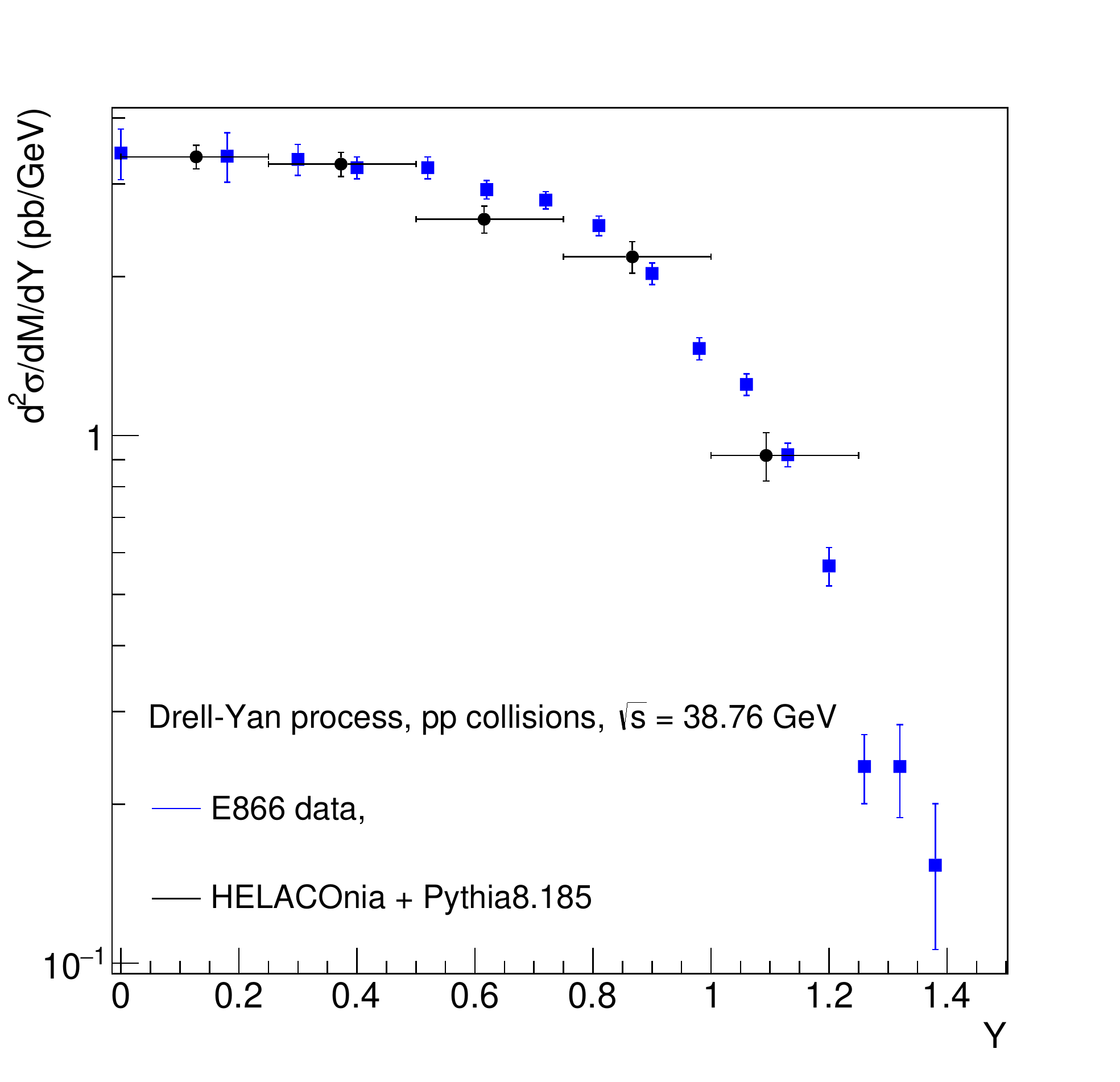}
		\caption{Drell-Yan cross section as a function of the rapidity in the center-of-mass frame obtained with \HELACOnia + \Pythia at $\sqrt{s} = 38.76$~GeV and rescaled by a factor 1.2, together with E866 data extracted from~\cite{Anastasiou:2003yy}. The invariant mass range considered is $7.2 < M < 8.7 $ GeV/$c^{2}$.}
		\label{fig:STARccbar}
\end{figure}

\subsection{Uncorrelated background}

The uncorrelated background was obtained from a minimum bias \Pythia $p+p$ simulation at $\sqrt{s} = 115$~GeV using the process \texttt{SoftQCD:nonDiffractive} with the MRSTMCal.LHgrid LHAPDF (6.1.4) set~\cite{Buckley:2014ana}. By comparing our simulation of open charm with a low statistic pure minimum bias \Pythia simulation, we have checked that the contribution of dimuons originating from a muon from charm/beauty and a muon from $\pi$/$K$ is negligible. The dominant source of uncorrelated opposite-sign muon pairs is the simultaneous semi-muonic decay of uncorrelated $\pi$ and/or $K$. In order to avoid possible double counting of signal and correlated background processes, the following hard processes have been switched off from the minimum bias simulations: \texttt{HardQCD:hardccbar}, \texttt{HardQCD:hardbbbar}, \texttt{WeakSingleBoson:ffbar2gmZ}\footnote{in order to avoid Drell-Yan pair production.}, \texttt{Charmonium:all} and \texttt{Bottomonium:all}, .

\section{Fast simulation of the response of a LHC\lowercase{b}-like detector}
\label{sec:fast}

The \HELACOnia and \Pythia generators provide the opposite-sign muon pairs from quarkonia decays, correlated and uncorrelated backgrounds sources, as defined in the previous section. In order to account for the detector resolution and the particle identification capabilities of a given detector and to investigate the feasibility of the quarkonium studies in $p+p$ collisions at $\sqrt{s} \simeq 115$~GeV, the detector response needs to be simulated.
For this purpose we have chosen a detector setup similar to the LHCb detector~\cite{Alves:2008zz}. The forward detector is very well suited as a fixed-target experiment setup as well, with a good tracking and particle identification capabilities.

\begin{figure*}[ht!]
		\centering
		\includegraphics[width=\columnwidth]{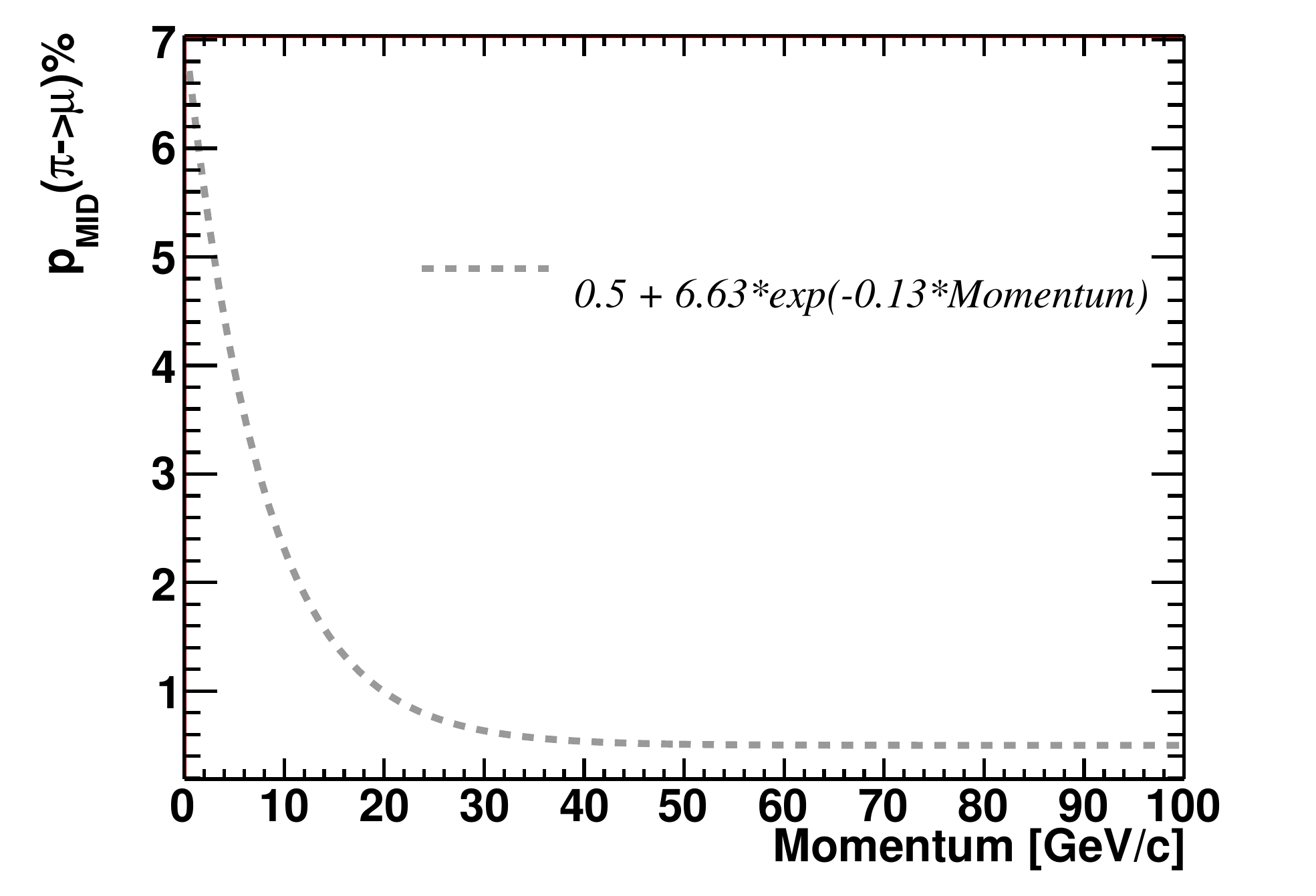}
		\includegraphics[width=\columnwidth]{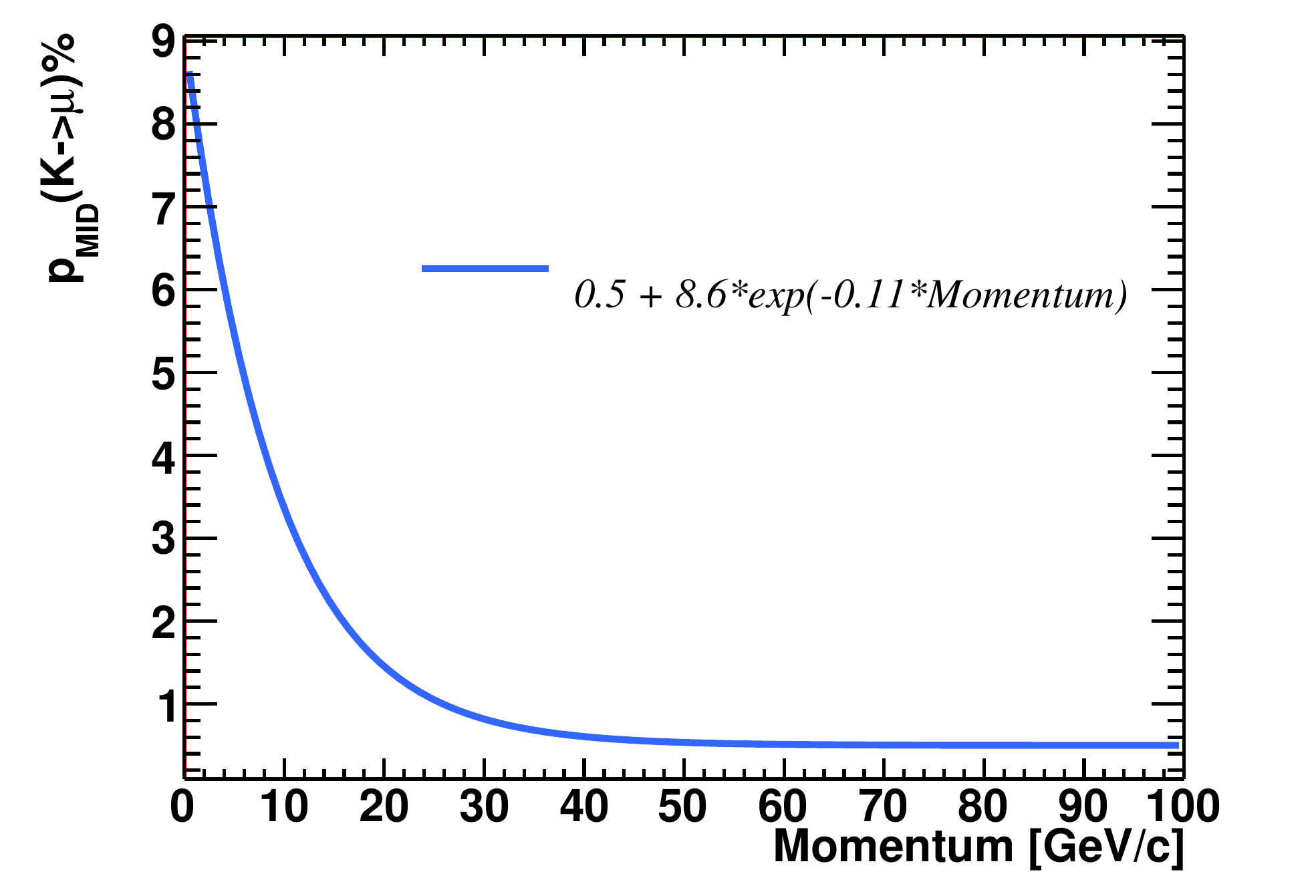}
		\caption{Misidentification probability of $\pi$ (left plot) and $K$ (right plot) as muon candidates as a function of momentum, $P_{MID}(\pi \rightarrow \mu)$ and $P_{MID}(K \rightarrow \mu)$, respectively.}
		\label{fig:muMisIdentification}
\end{figure*}

\begin{figure*}[ht!]
		\centering
		\includegraphics[width=0.65\columnwidth]{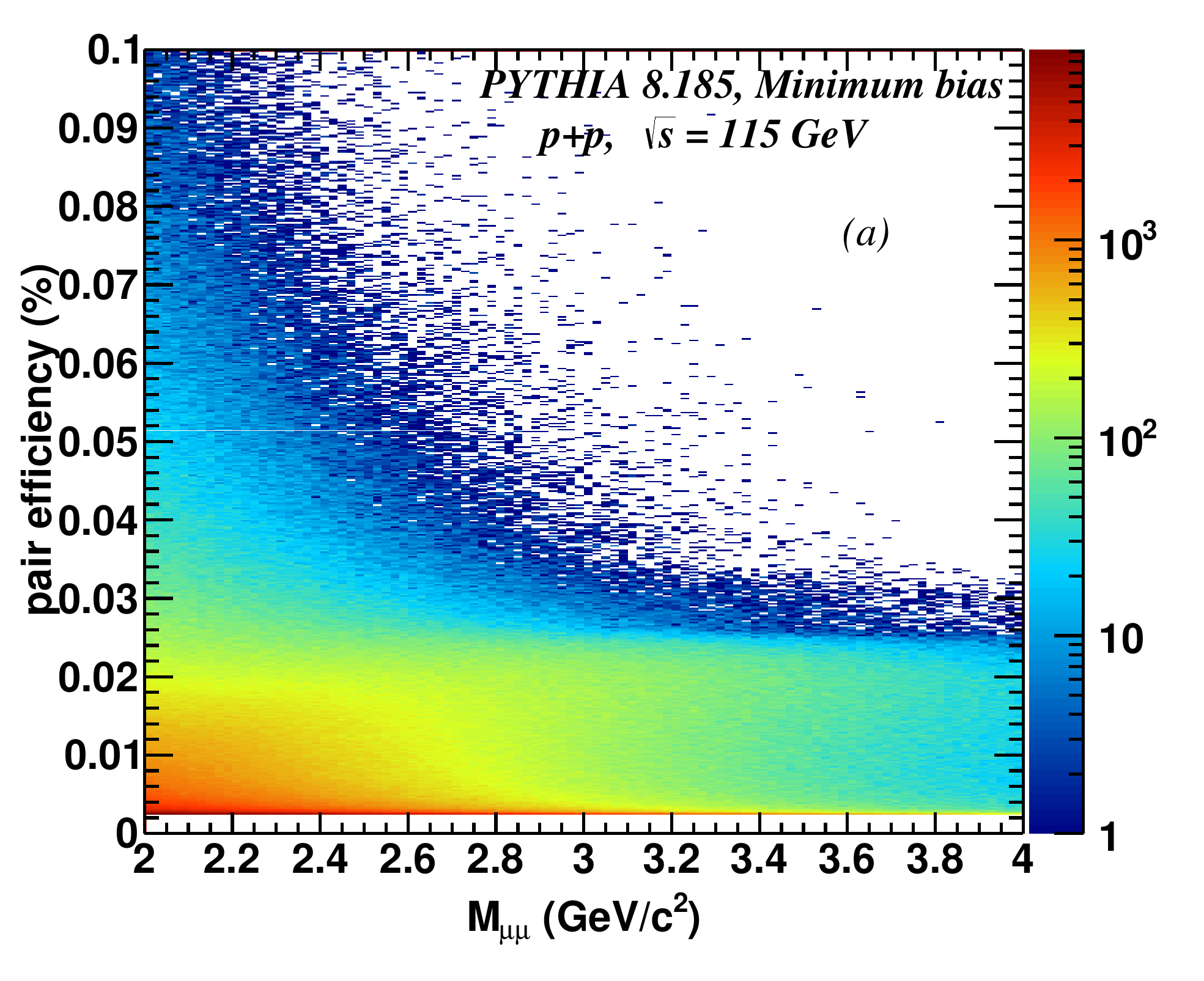}
		\includegraphics[width=0.65\columnwidth]{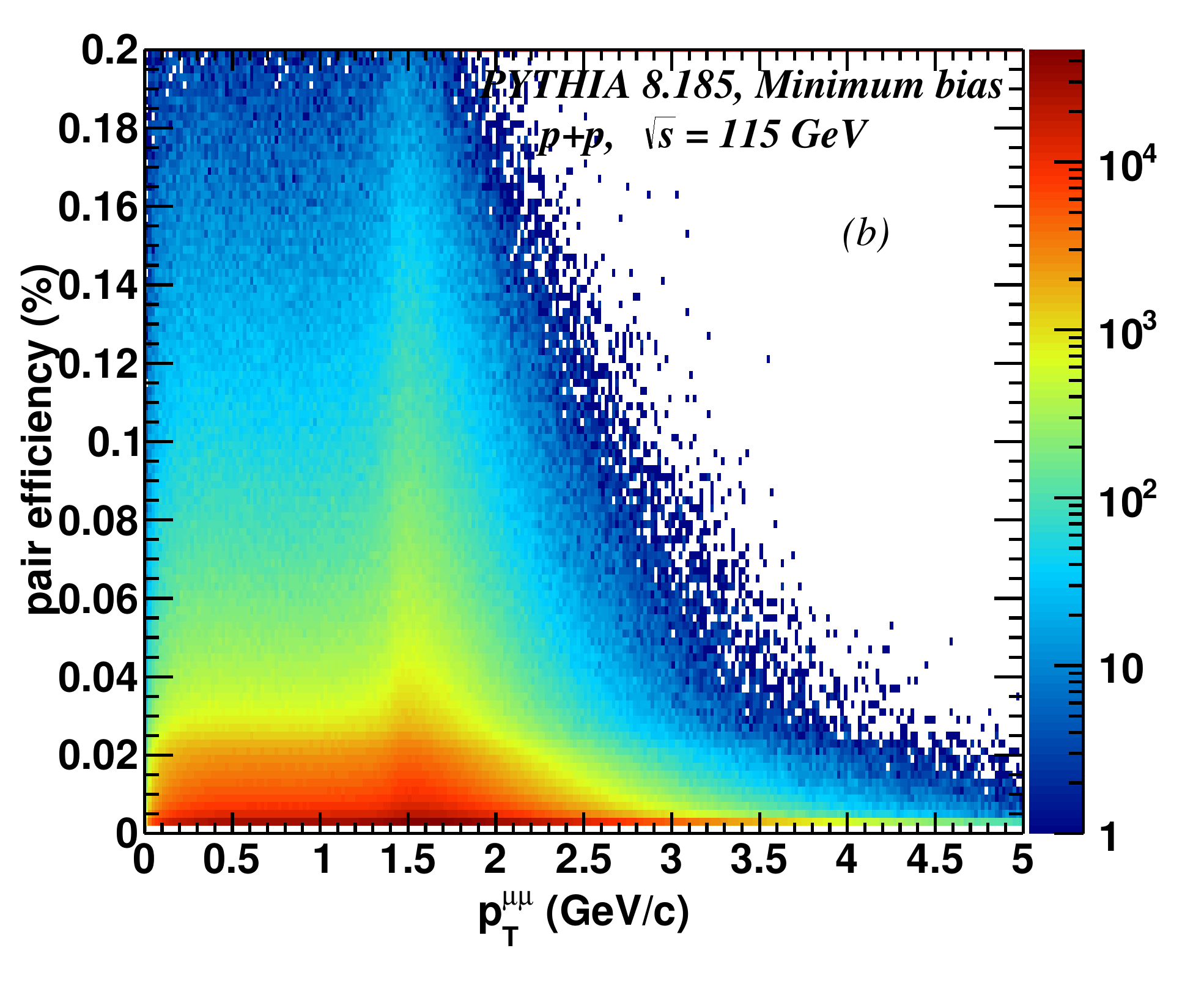}
		\includegraphics[width=0.65\columnwidth]{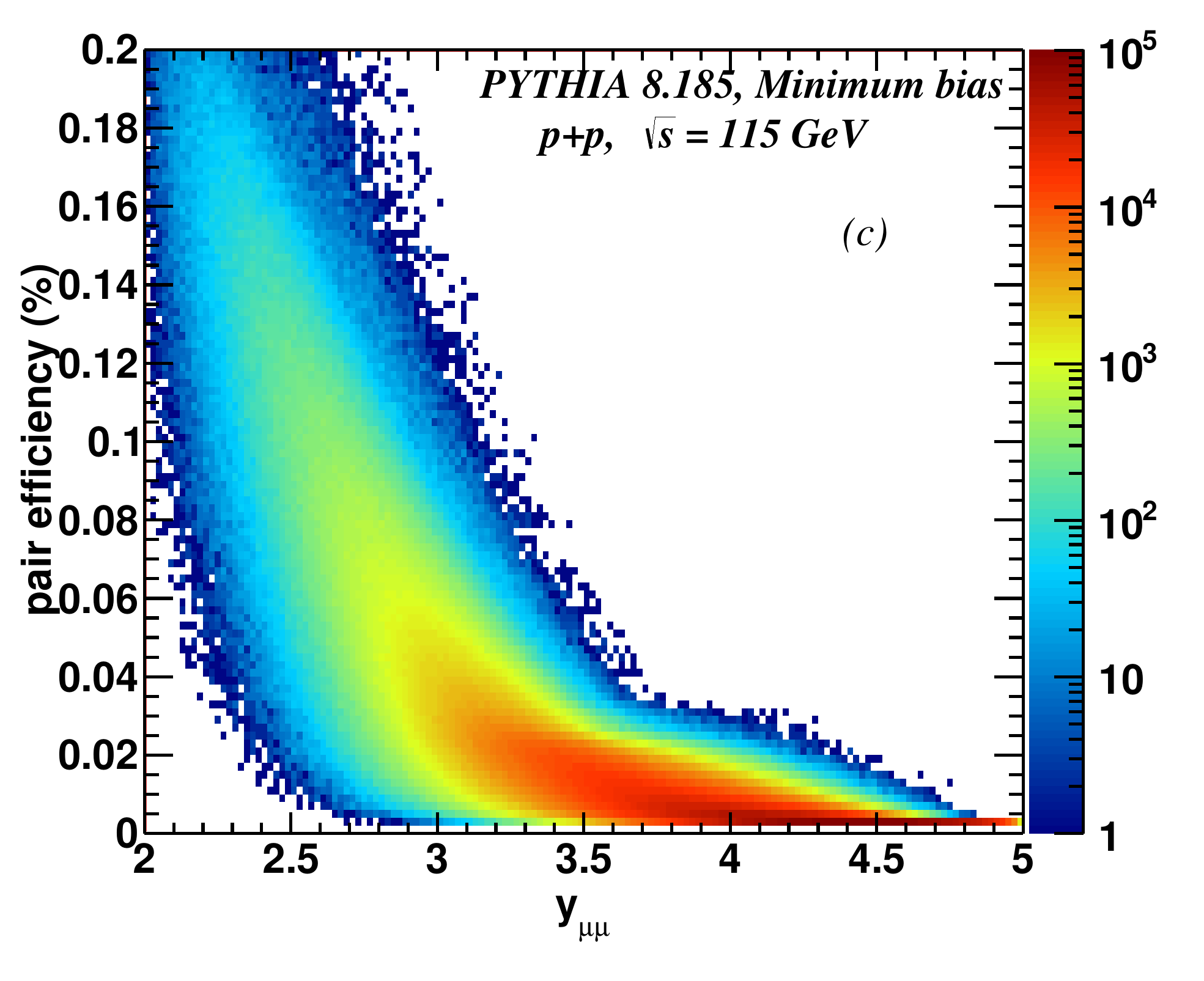}
		\caption{Muon pair, $\mu^{+}\mu^{-}$, identification efficiency as a function of the pair invariant mass (a), transverse momentum (b) and rapidity (c) for uncorrelated muon background. The efficiency takes into account the identification efficiency of the prompt muons, and the $\pi$  and $K$ misidentification probability , $P_{MID}(\pi \rightarrow \mu)$ and  $P_{MID}(K \rightarrow \mu)$.}
		\label{fig:muonPairEfficiency}
\end{figure*}

According to LHCb analysis cuts, muons in our simulations are required to have their transverse momentum satisfying 
$p_\mathrm{T} > 0.7$~GeV/$c$~\cite{Aaij:2011jh} and their pseudo-rapidity  in the laboratory frame satisfying $2 < \eta < 5$. 
The $\eta$ cut range corresponds to the LHCb detector coverage. Since the momentum resolution reported by LHCb 
is $\delta p /p \sim$~0.4~(0.6)\% for a momentum of 3~(100)~GeV/$c$~\cite{Archilli:2013npa} we consider a momentum 
resolution of $\delta p / p =$ 0.5 \%. The single $\mu$ identification efficiency is taken to be $\epsilon _{P} =$~98\%, 
which is an average efficiency obtained by LHCb for muons coming from J/$\psi$ decays, for $p > 3$~GeV/$c$ and 
$p_\mathrm{T} > 0.8$~GeV/$c$ \cite{Archilli:2013npa}. These cuts and the abovementioned detector response on the muons are applied to 
simulate the quarkonium states and all the background sources.

In the case of uncorrelated background, as discussed in section~\ref{sec:input}, most of the $\mu$  originate 
from $\pi^{+/-}$ or $K^{+/-}$ decays. If a $\pi$ or $K$ decays to a $\mu$ before 12~m along the z axis, the $\mu$ 
is rejected by the tracking system and it is not considered in the simulation. 12 m corresponds to the distance 
where calorimeters, followed by muon stations, are placed in the LHCb detector setup. If the $\mu$ is produced beyond 12~m 
or if a $\pi/K$ is misidentified with $\mu$ in the muon stations, a $\pi/K$ misidentification probability is applied. The 
misidentification probabilities depend on the total particle momentum and were reported by the LHCb collaboration 
in~\cite{Aaij:2014jba}. These probabilities are parametrised with the following functions: $P_{MID}(\pi \rightarrow \mu)(p) = (0.5 + 6.63 \exp(-0.13p)) \%$ and $P_{MID}(K \rightarrow \mu)(p) = (0.5 + 8.6 \exp(-0.11p)) \%$, and are shown in \cf{fig:muMisIdentification} (a) and (b), for $\pi$ and K respectively.
Based on the single $\mu$ identification efficiency $\epsilon _{\mu^{+/-}}$, the dimuon, $\mu^{+} \mu^{-}$, efficiency is calculated as a product of the single efficiencies: $\epsilon _{\mu^{+}\mu^{-}} = \epsilon_{\mu^{+}} \times \epsilon_{\mu^{-}}$. For muons coming from $\pi^{+\-}$ or $K^{+/-}$ decays, misidentification probabilities are used: $\epsilon_{\mu^{+/-}} = P_{MID}(\pi \rightarrow \mu)(p)$ or  $\epsilon_{\mu^{+/-}} = P_{MID}(K \rightarrow \mu)(p)$,  respectively for $\pi$ and $K$, and for prompt muons $\epsilon_{\mu^{+/-}} = \epsilon _{P} = 0.98$.

The pair efficiency is extracted in each kinematic phase-space point 
and is shown as a function of the dimuon invariant mass, transverse momentum and rapidity 
in \cf{fig:muonPairEfficiency}. This efficiency is used to correct dimuon spectra 
obtained with the uncorrelated background \Pythia simulations.

\section{Quarkonium production studies in \lowercase{p}H collisions at $\sqrt{s}=115$~GeV}
\label{sec:simupp}

\begin{figure*}[ht]
		\centering
		\includegraphics[width=0.9\columnwidth]{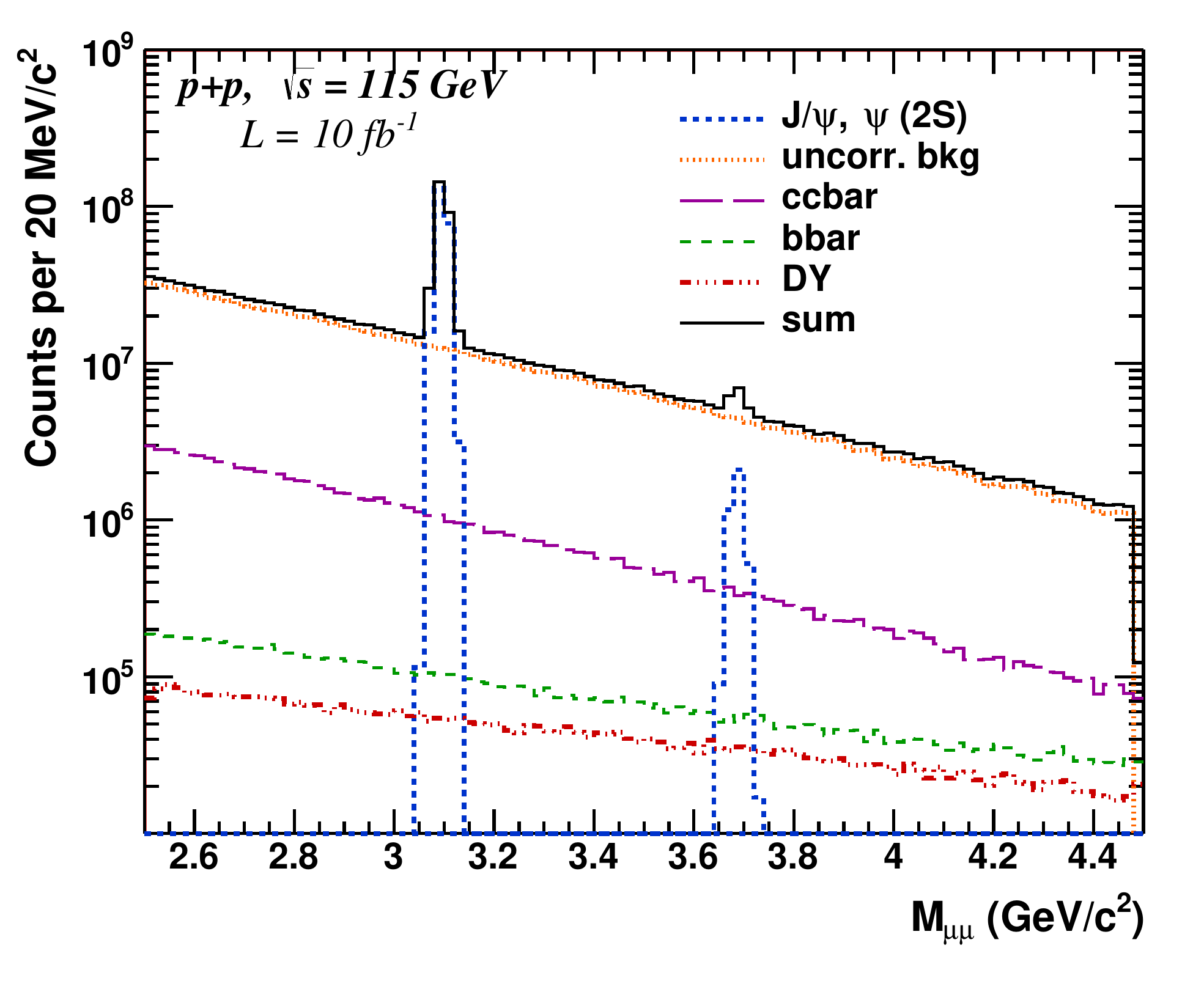}
		\includegraphics[width=0.9\columnwidth]{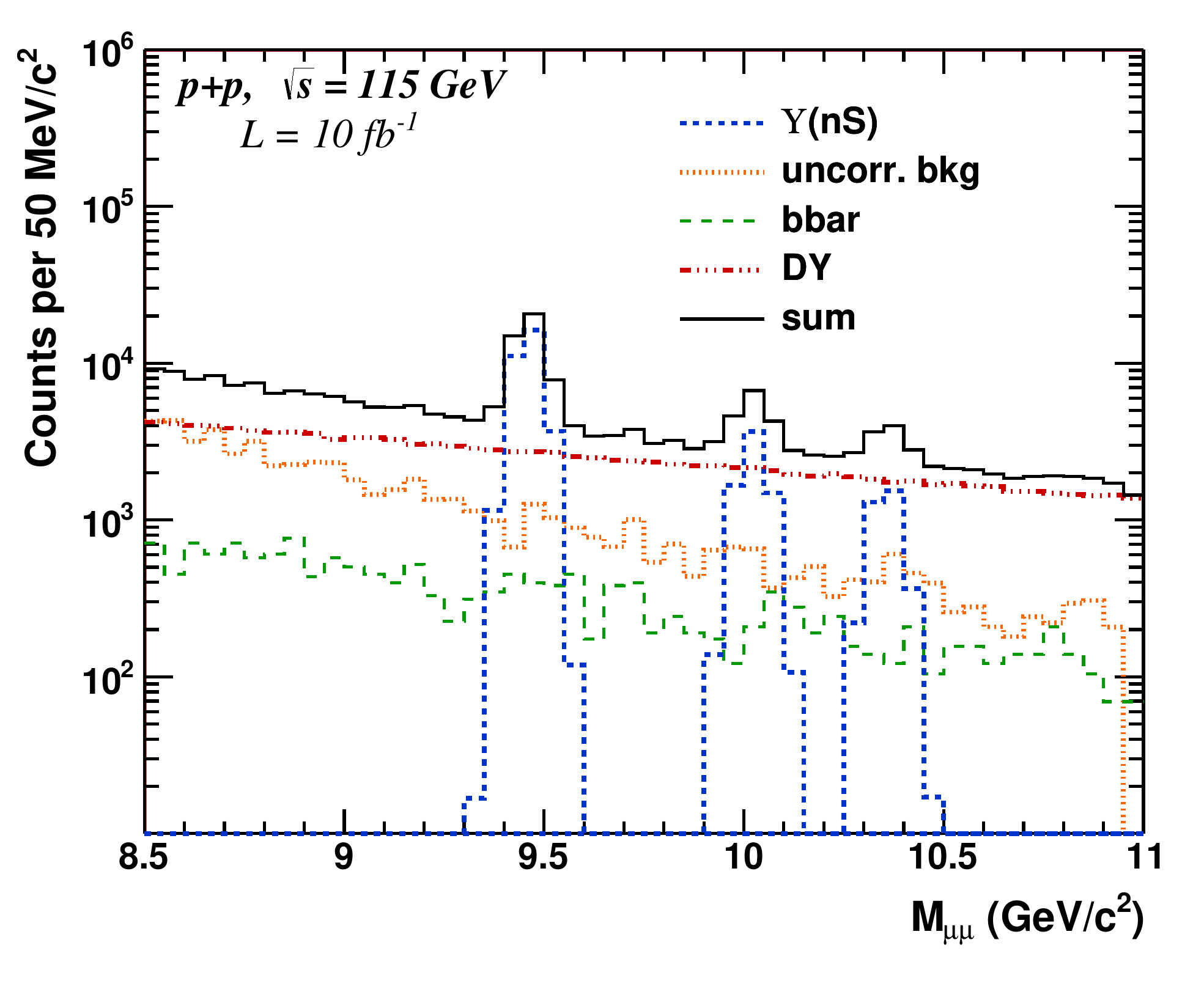}
		\caption{Dimuon invariant mass distributions for J/$\psi$ and $\psi(2S)$, left plot, and $\Upsilon(nS)$, right plot, with different background sources. }
		\label{fig:QuarkoniaWithBkg}
\end{figure*}

\begin{figure*}[ht]
		\centering
		\includegraphics[width=0.9\columnwidth]{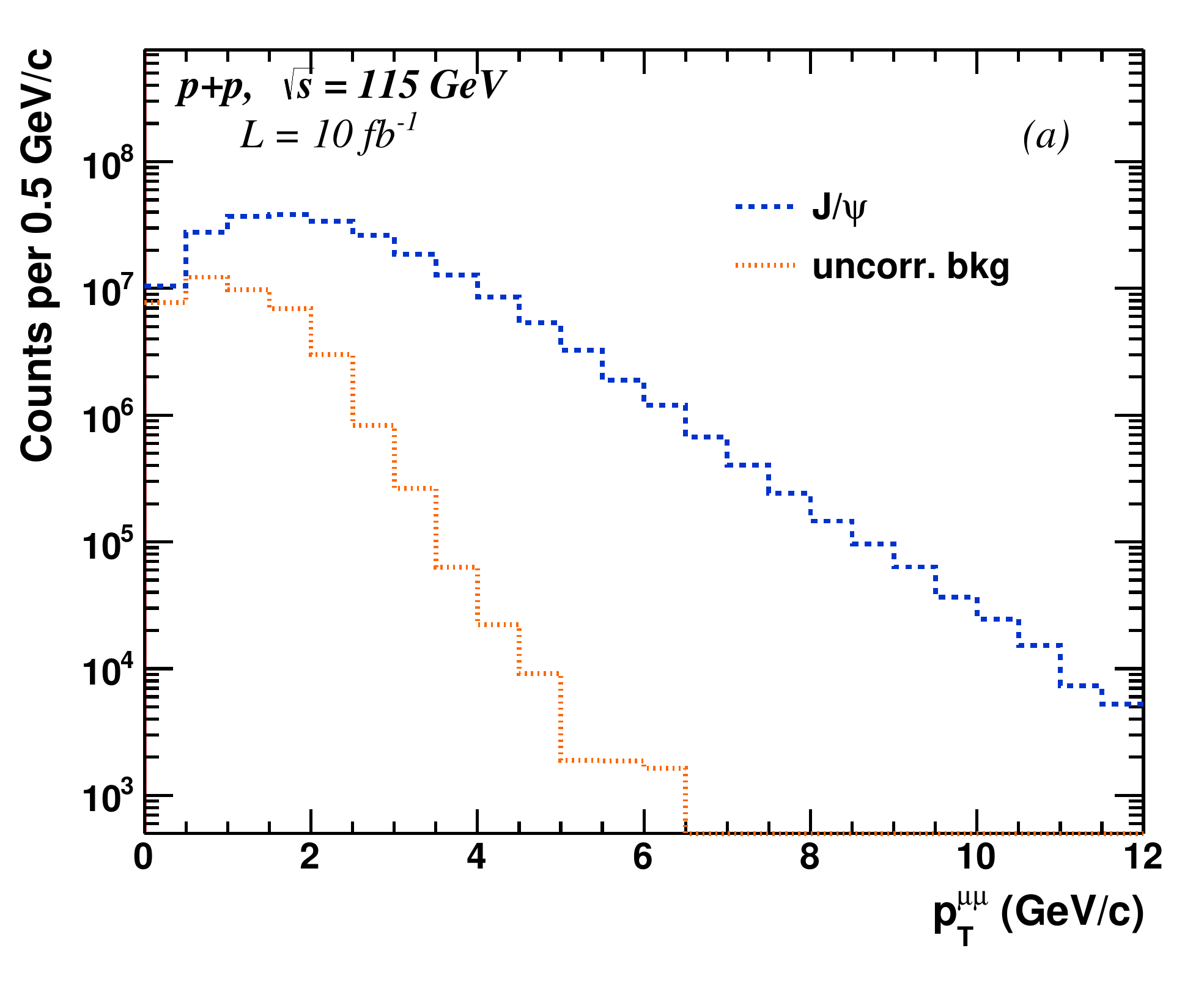}
		\includegraphics[width=0.9\columnwidth]{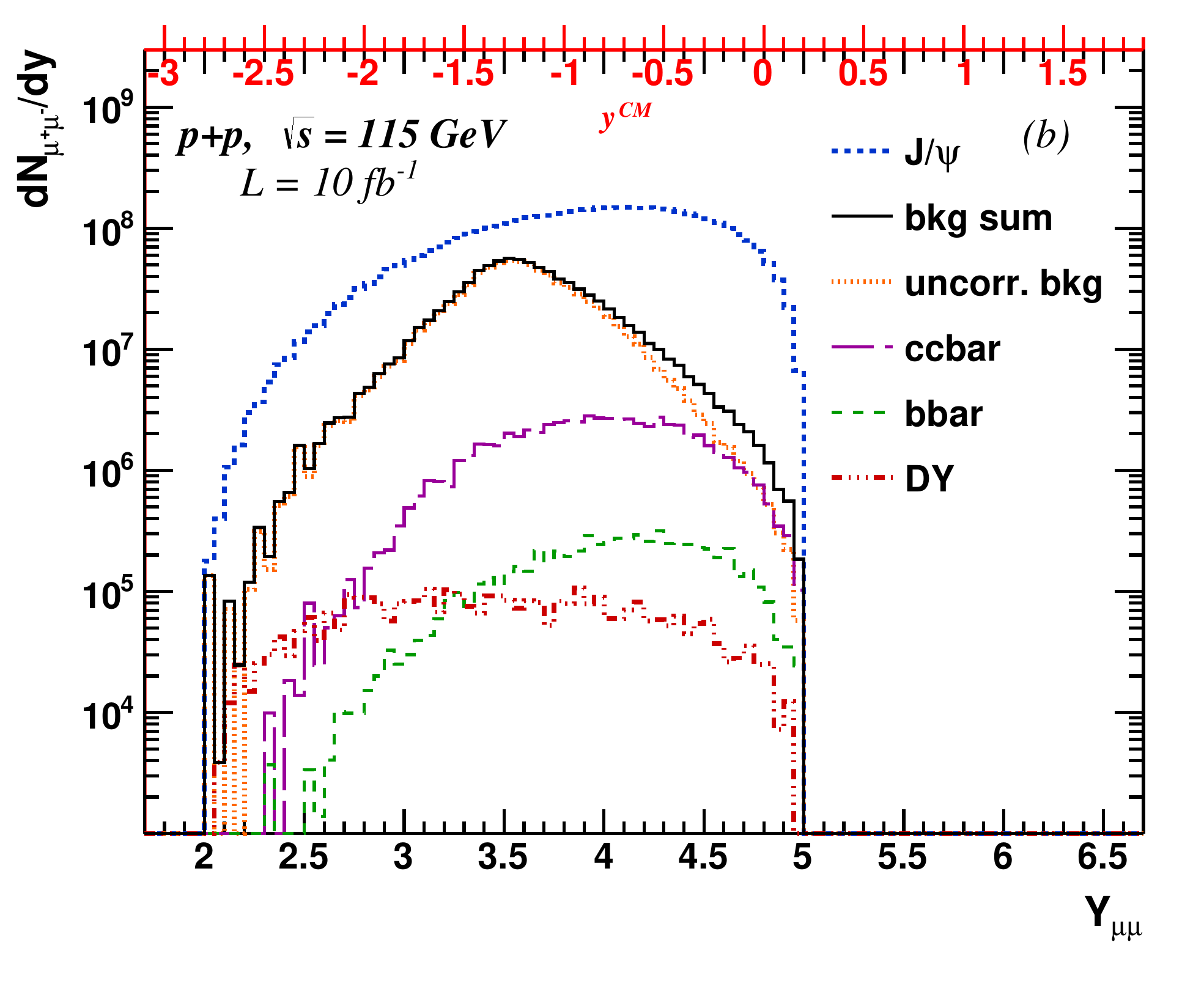}
		\caption{$p_\mathrm{T}$ (a) and y (b) spectra of J/$\psi$ signal and different background sources, in the J/$\psi$ mass range.}
		\label{fig:JpsiPtYWithBkg}
\end{figure*}

In this section, we show results on the quarkonium production studies in the dimuon decay channels, with the dominant background sources. Simulations have been performed for a 7 TeV proton beam on a hydrogen target ($p+p$), which gives $\sqrt{s} = 115$~GeV. 
We consider an integrated luminosity of 10~fb$^{-1}$ which is expected to be obtained after half of a LHC year 
with the crystal mode, as described in section~\ref{sec:intro} and \ct{tablumi}.

\begin{figure*}[ht]
		\centering
		\includegraphics[width=.675\columnwidth]{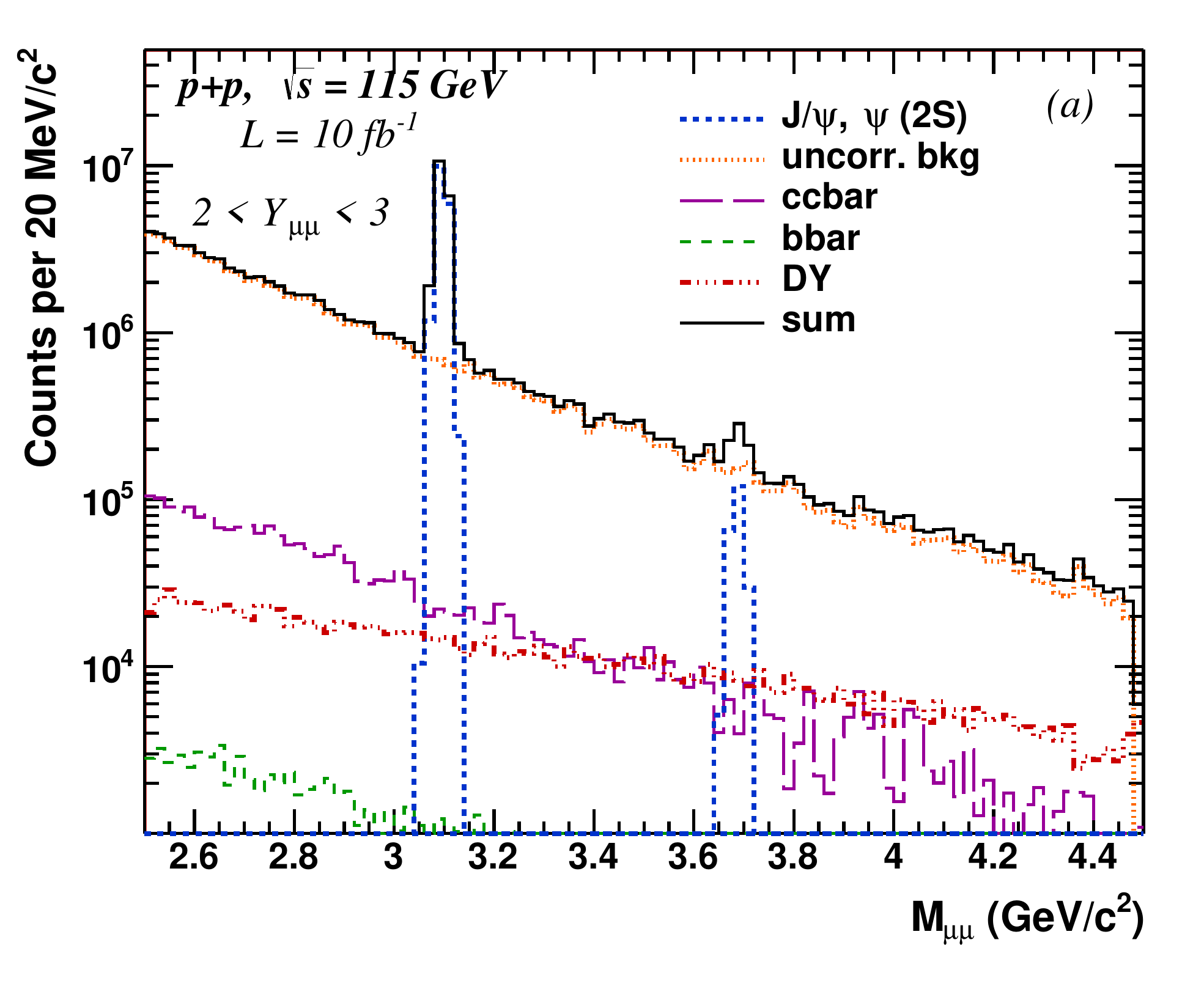}
		\includegraphics[width=.675\columnwidth]{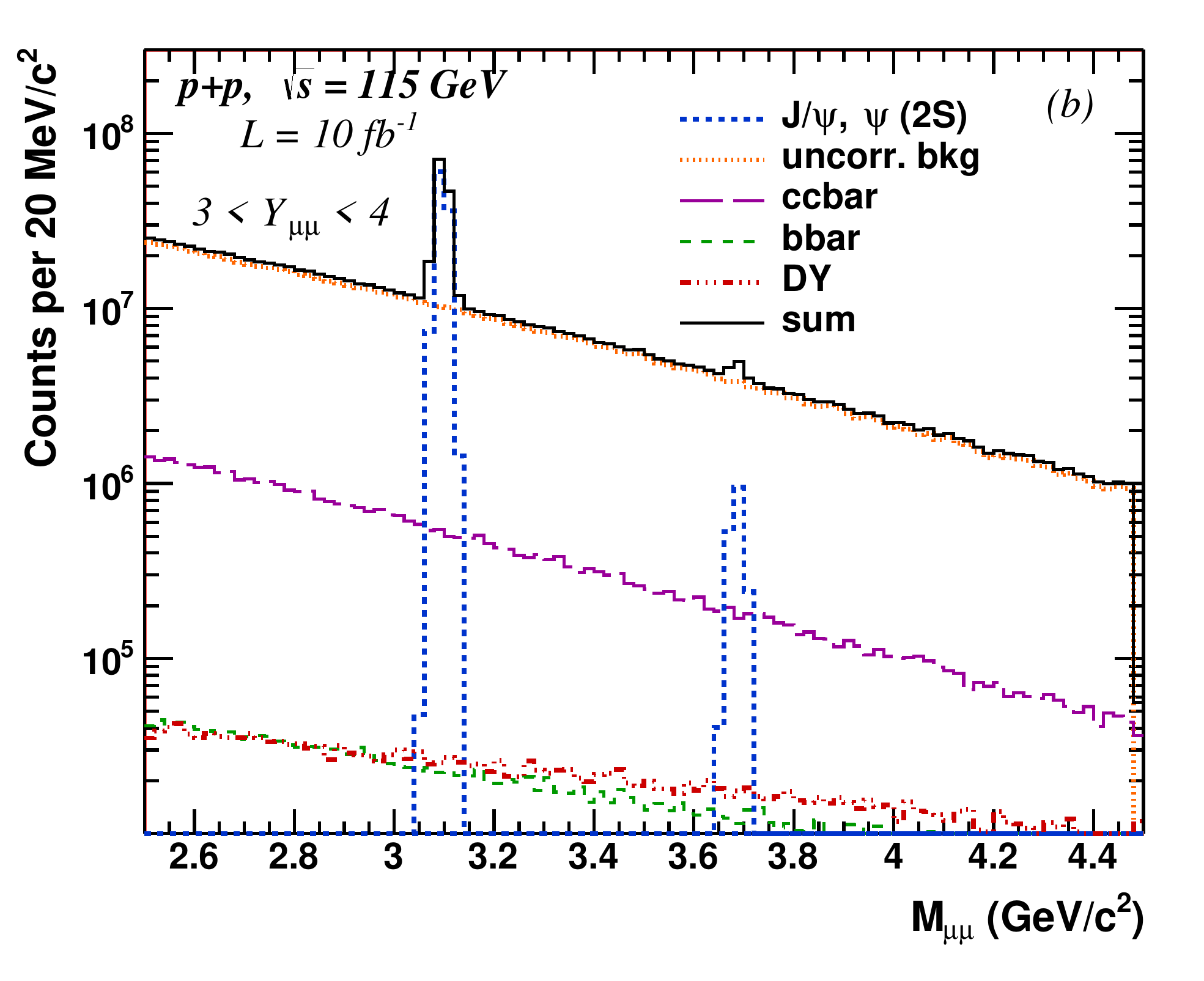}
		\includegraphics[width=.675\columnwidth]{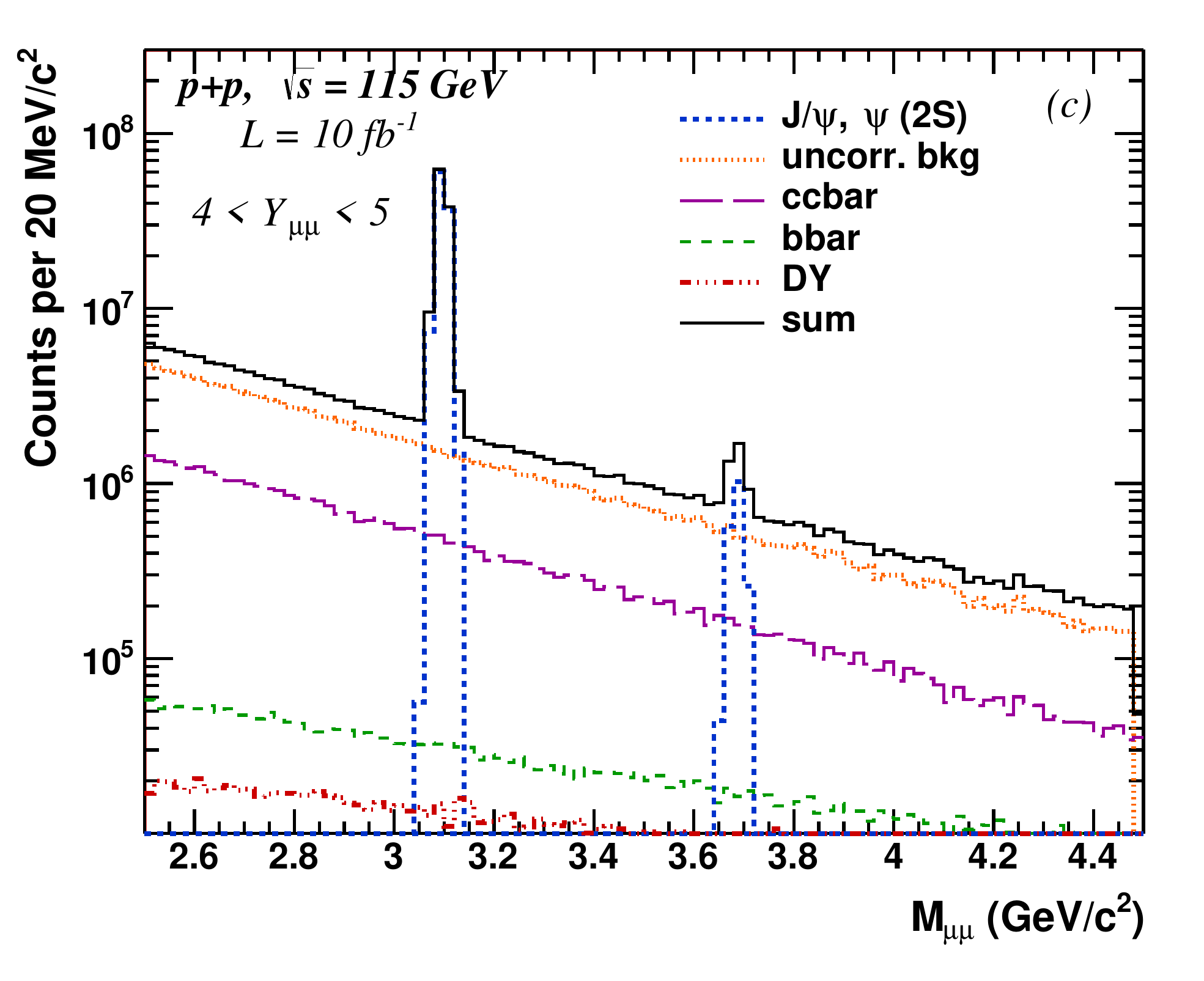}
		\caption{Dimuon invariant mass distributions, for three rapidity bins, in J/$\psi$ and $\psi(2S)$ mass window. $2 < Y_{\mu^{+}\mu^{-}} < 3$, $3 < Y_{\mu^{+}\mu^{-}} < 4$, $4 < Y_{\mu^{+}\mu^{-}} < 5$, shown on panels (a), (b) and (c), respectively. }
		\label{fig:QuarkoniaWithBkg_ybins}
\end{figure*}

\subsection{Background studies}

These simulations allow to quantify the background sources in the quarkonium studies in the dimuon decay channel, which could potentially make the quarkonium signal extraction more difficult or even prevent from obtaining a clear signal. In particular, this may be critical for the excited states.
We present here simulations of invariant mass of opposite-sign muon pairs, $\mu^{+}\mu^{-}$, from the quarkonia and from the dominant background sources, in two mass ranges, see \cf{fig:QuarkoniaWithBkg}. The first range corresponds to the J/$\psi$ and $\psi(2S)$ invariant mass windows and the second one to the mass range of the $\Upsilon(1S)$, $\Upsilon(2S)$ and $\Upsilon(3S)$. The invariant mass distributions are integrated over the whole transverse momentum and rapidity ranges. The plots show the simulated quarkonium signals and the background, separately from the different sources, and the black solid line is a sum of all contributions. The background sources correspond to an integrated luminosity of 10~fb$^{-1}$.

\begin{figure*}[ht]
		\centering
		\includegraphics[width=0.9\columnwidth]{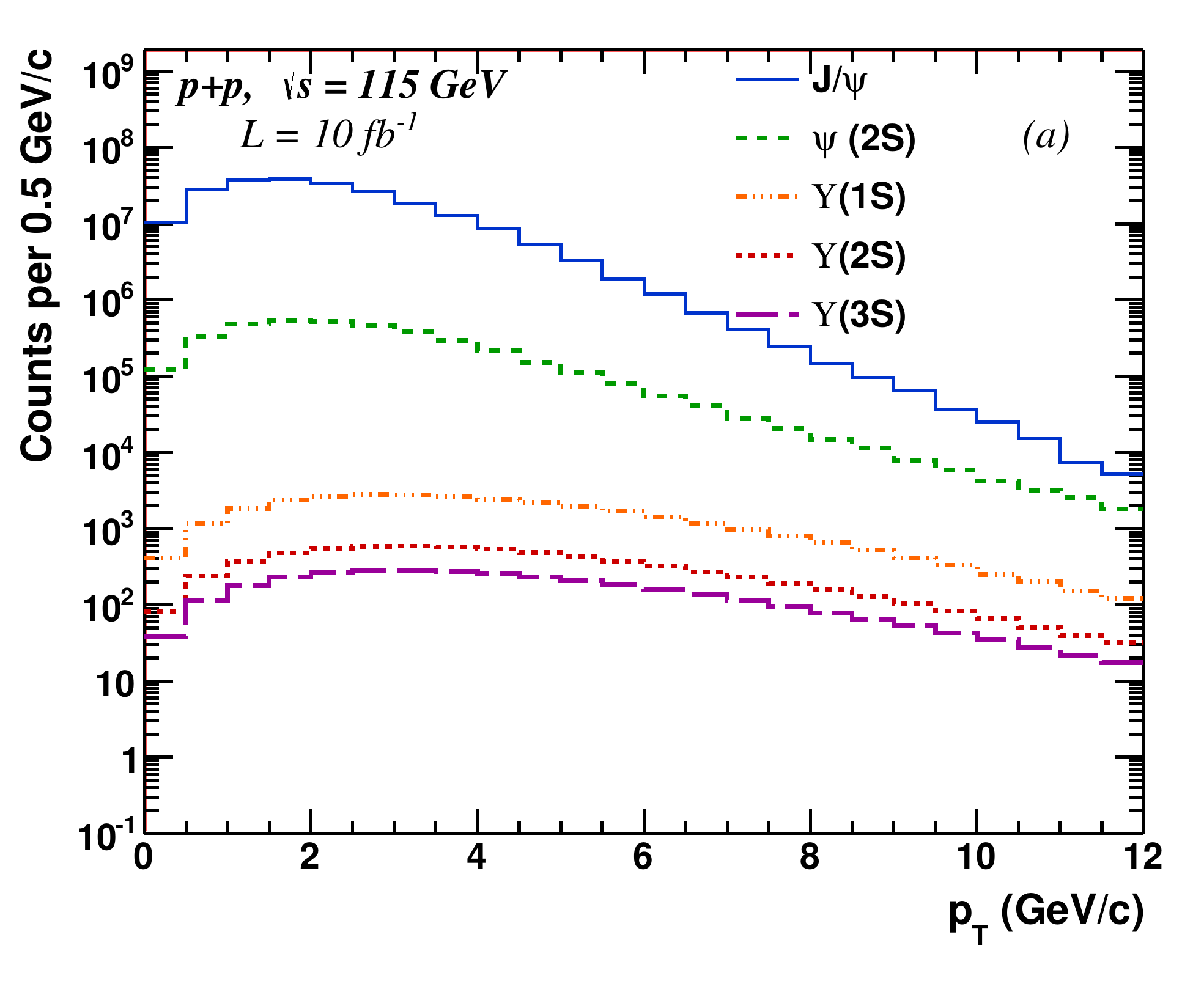}
		\includegraphics[width=0.9\columnwidth]{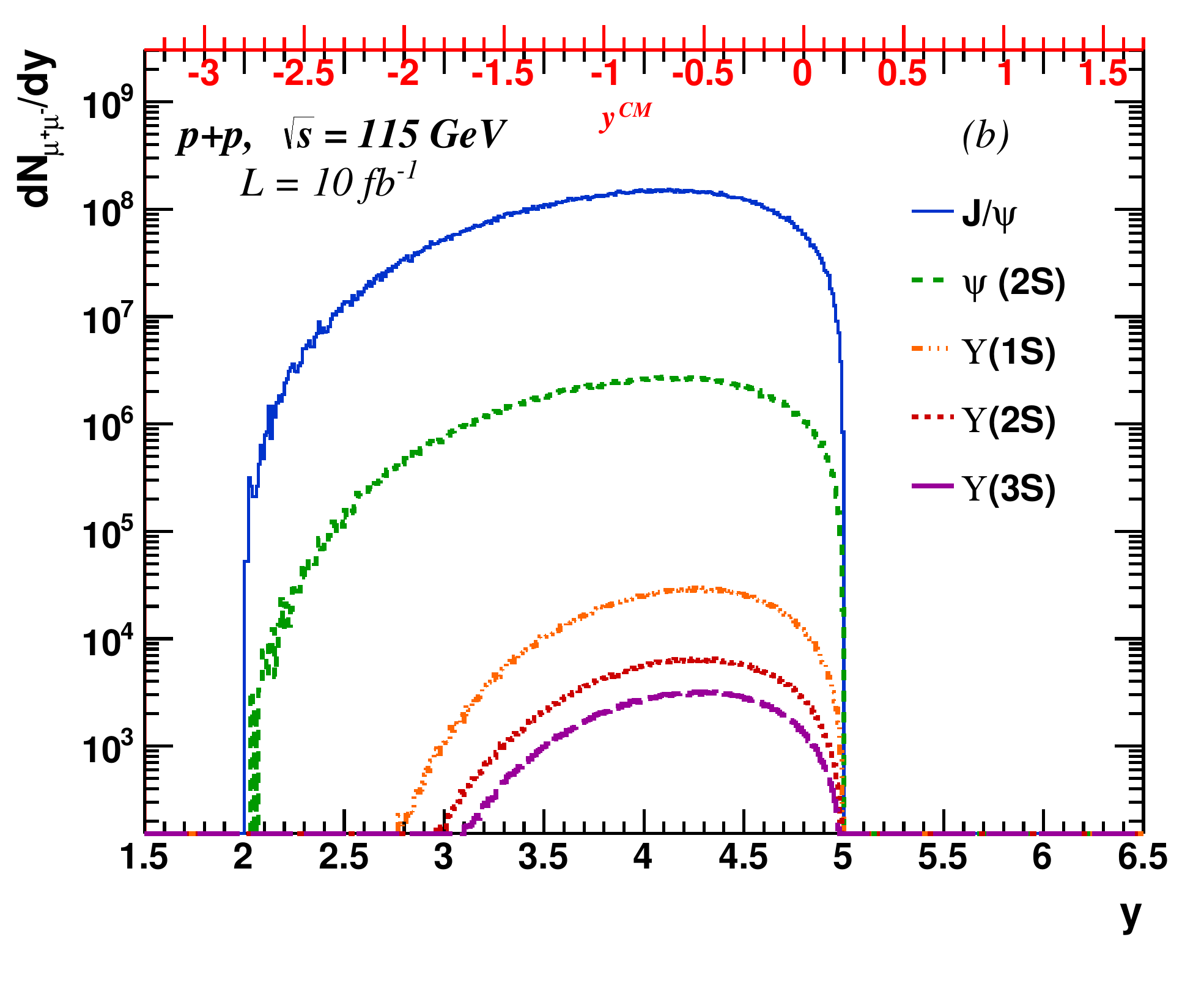}
		\caption{Transverse momentum (a) and rapidity (b) distributions for J/$\psi$, $\psi(2S)$, $\Upsilon(1S)$, $\Upsilon(2S)$ and $\Upsilon(3S)$, from the top to the bottom distribution.}
		\label{fig:QuarkoniaPtY}
\end{figure*}

In the J/$\psi$ and $\psi(2S)$ invariant mass window, the dominant background source is from uncorrelated $\mu^{+}\mu^{-}$ pairs, mostly from $\pi^{+/-}$ and $K^{+/-}$ decays. Contributions from Drell-Yan and $b\bar{b}$ continuum are very small. In the case of $\Upsilon(nS)$ states the Drell-Yan contribution is the dominant one. Under the $\Upsilon(nS)$ peak, a contribution from the $c\bar{c}$ continuum is negligible, and is not considered here.

The significance ($sig = S/\sqrt{(S+B)}$, where S is the number of signal counts and B the number of background counts, in the invariant mass range $M_{Q} \pm 3\, \sigma_{Q}$) and the signal to background ratio ($S/B$) of each quarkonium state are given in the following:
\begin{itemize}
\item $sig_{J/\psi} =$ 134.6 $10^{2}$ $\sigma$, $S/B _{J/\psi} =$ 4.21
\item $sig_{\psi (2S)} =$ 735.2  $\sigma$, $S/B _{\psi (2S)} =$ 0.16
\item $sig_{\Upsilon (1S)} =$ 140.73 $\sigma$, $S/B _{\Upsilon (1S)} =$ 1.75
\item $sig_{\Upsilon (2S)} =$ 45.29 $\sigma$, $S/B _{\Upsilon (2S)} =$ 0.48
\item $sig_{\Upsilon (3S)} =$ 25.75 $\sigma$, $S/B _{\Upsilon (3S)} =$ 0.28
\end{itemize}
for J/$\psi$, $\psi(2S)$, $\Upsilon(1S)$, $\Upsilon(2S)$ and $\Upsilon(3S)$, respectively.

Transverse momentum and rapidity distributions for the quarkonium signals and for each background source were also studied. As an example, $p_\mathrm{T}$ and y distributions in the J/$\psi$ mass range, 3.063 $< M_{\mu^{+}\mu^{-}} <$ 3.129 GeV/$c^{2}$ (corresponding to $M_{J/\psi} \pm 3~\sigma_{J/\psi}$) are shown in \cf{fig:JpsiPtYWithBkg}. It is visible that the distributions for the J/$\psi$ and different backgrounds differ. In more backward or forward rapidity regions, the signal to background ratio increases. This can also be seen in \cf{fig:QuarkoniaWithBkg_ybins}, where the dimuon invariant mass distributions in J/$\psi$ and $\psi(2S)$ mass window are shown in three rapidity ranges. In terms of transverse momentum, one can obtain a very clean signal when going to higher $p_\mathrm{T}$. Above $\sim$ 4 GeV/$c$, the uncorrelated background starts to vanish. Since $c\bar{c}$, $b\bar{b}$ and Drell-Yan simulations are LO simulations, the $p_\mathrm{T}$ spectra of these correlated background sources are not shown here.

\subsection{Quarkonium simulations}

We have also studied the $p_\mathrm{T}$ and rapidity coverage reach of the quarkonium signals.
The transverse momentum distributions are shown in \cf{fig:QuarkoniaPtY} (a), for J/$\psi$, $\psi(2S)$, $\Upsilon(1S)$, $\Upsilon(2S)$ and $\Upsilon(3S)$, from the top to the bottom distribution. Similarly, \cf{fig:QuarkoniaPtY} (b) shows the rapidity distribution for each quarkonium state. 
With an integrated luminosity of 10 fb$^{-1}$ the quarkonium studies can be carried out in a wide rapidity and $p_\mathrm{T}$ range. It should be possible to study $\Upsilon (nS)$ signals up to $p_\mathrm{T} \simeq 10$~GeV/$c$, and J/$\psi$ and $\psi (2S)$ could be studied even up to $p_\mathrm{T} \simeq$ 15 GeV/$c$. All the quarkonium states can be measured down to $p_\mathrm{T} = 0 $~GeV/$c$.

This study is limited by the rapidity range of 2 $< y < $ 5, in the laboratory frame, due to the pseudorapidity cuts on the decay $\mu$. The red x-axis on the top of \cf{fig:QuarkoniaPtY} (b) denotes the rapidity in the center-of-mass frame. The rapidity shift for a 7 TeV proton beam on a fixed target is -4.8, \ie\ $y_{CM} = 0 \rightarrow y_{lab} = 4.8$. J/$\psi$ and $\psi(2S)$ signals can be studied in the whole mentioned rapidity range, while the lowest rapidity reach for $\Upsilon(nS)$ is $\sim$ 2.5 -- 3.

\section{Quarkonium measurements in \lowercase{p}+A collisions at $\sqrt{s}=115$~GeV and P\lowercase{b}+H collisions at $\sqrt{s}=72$~GeV }
\label{sec:simupA}

\subsection{Multiplicity in proton-nucleus collisions}

In proton-nucleus collisions, the high track multiplicity may induce a high detector 
occupancy and lead to a reduction of the detector capabilities. Since LHCb has successfully measured the 
\jpsi and \ups production 
in $p$+Pb collisions at $\snn=5$~TeV~\cite{Aaij:2013zxa, Aaij:2014mza}, one would expect a good capability of 
such detector under similar 
particle multiplicity environment. In the following, the charged particle multiplicity has been generated with 
the EPOS generator~\cite{Pierog:2013ria,Werner:2005jf} in different configurations: $p+\mathrm{Pb}$ collisions 
at $\snn=5$~TeV in collider mode, $p+\mathrm{Pb}$ collisions at $\snn=115$~GeV and $\mathrm{Pb}+\mathrm{H}$ 
collisions at $\snn=72$~GeV in fixed-target mode. 
The charged-particle multiplicity is dominated by the $\pi$ multiplicity. By comparing these three distributions 
as a function of the pseudorapidity of the particle in the laboratory frame as shown in \cf{fig:mult-pA}, 
one can conclude that the charged particle multiplicity 
in a fixed-target mode never exceeds the one obtained in $p+\mathrm{Pb}$ collisions at $\snn=5$~TeV in the collider 
mode in the full pseudorapidity range: a detector with the LHCb capabilities will be able to run in such 
conditions.

\begin{figure}[ht]
		\centering
		\includegraphics[width=0.8\columnwidth]{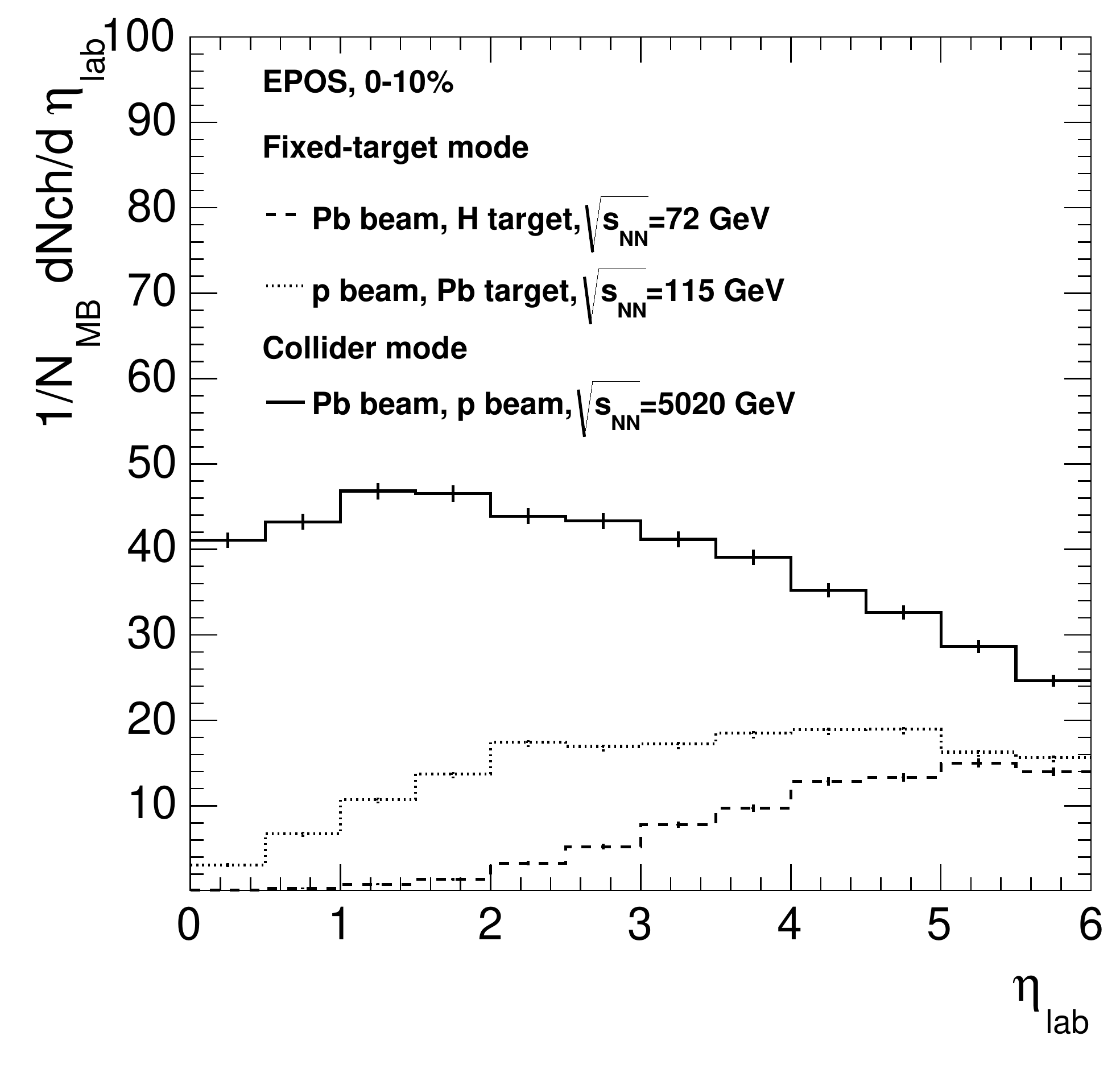}
		\caption{Averaged number of charged particles in $p+A$  collisions as a function of the pseudorapidity in the laboratory frame.}
		\label{fig:mult-pA}
\end{figure}

\subsection{Prospects for the measurements of the nuclear modification factors for J$/\psi$ and
$\Upsilon$ in  \lowercase{p}+Pb  collisions at $\snn=115$~GeV}

To illustrate the potential offered by AFTER@LHC in $p+\mathrm{Pb}$ collisions at $\snn=115$~GeV, we have evaluated, 
in this section, the impact 
of the nuclear modification of the gluon densities in nucleons within large nucleus --generically referred to as
gluon shadowing-- and its uncertainty as encoded in the nuclear PDF set EPS09. For that, we have used  
the probabilistic Glauber Monte-Carlo framework, {\sf JIN}~\cite{Ferreiro:2008qj,Ferreiro:2008wc},
which allows us to encode different mechanisms for the partonic production and to interface these production processes
with different cold nuclear matter effects, such as the aforementioned shadowing, in order to get the production cross sections for
proton-nucleus and nucleus-nucleus collisions.  {\sf JIN} also straightforwardly 
computes any nuclear modification factor for minimum bias collisions or
in specific centrality classes. In the case of proton-nucleus ($p+A$) collisions, it is 
the ratio of the yield per inelastic collision in $p+A$ collisions to the yield in
$pp$ collisions at the same energy multiplied by the average
number of binary collisions in a typical $p+p$ collision, $\langle \Ncoll\rangle$:
\begin{equation}
R_{pA}=\frac{dN_{pA}^{}}{\langle\Ncoll\rangle dN_{pp}^{}}.
\end{equation}
In the presence of a net nuclear effect,  $R_{pA}$ is defined such that it differs from {\it unity}.
In the simplest case of minimum bias collisions, one should have
\begin{equation}
R_{pA} =\frac{d\sigma_{pA}^{}}{A d\sigma_{pp}^{}}.
\end{equation}

As in \cite{Ferreiro:2013pua}, we have used the central curve of EPS09 as well as 
four specific extreme curves (minimal/maximal shadowing, minimal/maximal EMC effect), 
which reproduce the envelope of the gluon nPDF uncertainty encoded in EPS09 LO~\cite{Eskola:2009uj}.

In addition to the modification of the partonic densities, quarkonium production
in $p+A$ collisions can be affected by other effects, 
for instance by the nuclear absorption which depends much 
on the nature of the object traversing the nuclear medium. If the meson is already formed, it may be 
affected more than a smaller pre-resonant pair. To discuss such an effect, it is useful to introduce the 
concept of the formation time, $t_f$, based on the Heisenberg uncertainty principle and the time 
--in the rest frame of the meson-- to discriminate between two $S$ states, for instance
the J$/\psi$ and the $\psi(2S)$. In fact, one finds~\cite{Ferreiro:2013pua,Ferreiro:2011xy} that such a time is similar 
for the charmonium and bottomonium states and is on the order of 0.3-0.4 fm. Obviously, this time has to be boosted
in the frame where the nuclear matter sits. For $t_f$ smaller than the nucleus radius, the quarkonium is formed before escaping
it. In the fixed-target mode with a proton beam and a nuclear target, 
the boost factor is simply $\gamma(y_{lab})=\cosh(y_{lab})$. We therefore obtain $t_f$ as in \ct{tab:tf-AFTER}.

\begin{table}[htb!]
\begin{center}
\begin{tabular}{ccccc|ccccc}
\hline
 $y_{\rm CMS}$ & $y_{lab}$ &$\gamma(y_{lab})$ & $t^{\mathrm{J}/\psi,\Upsilon}_f(y)$   & \quad& & $y_{\rm CMS}$ & $y_{lab}$ & $\gamma(y_{lab})$ & $t^{\mathrm{J}/\psi,\Upsilon}_f(y)$\\
\hline
-2.5 & 2.3  &   5 & 1.75 fm &   &   &  -0.5 & 4.3 &37      & 13 fm\\
-1.5 & 3.3  &  14 & 5 fm &   &   & 0. & 4.8 &61      & 21 fm\\
-1.0 & 3.8  &  22  & 8 fm &   &   & 0.5 & 5.3 &100      & 35 fm\\
\hline
\end{tabular}
\caption{Boost and formation time in the (target) Pb rest frame of a J$/\psi$ and a $\Upsilon$ as a function of its CMS rapidity
at $\sqrt{s_{NN}}=115$ GeV.}\label{tab:tf-AFTER}
\end{center}\vspace*{-0.5cm}
\end{table}

\begin{figure*}[ht]
		\centering
		\includegraphics[width=0.9\columnwidth]{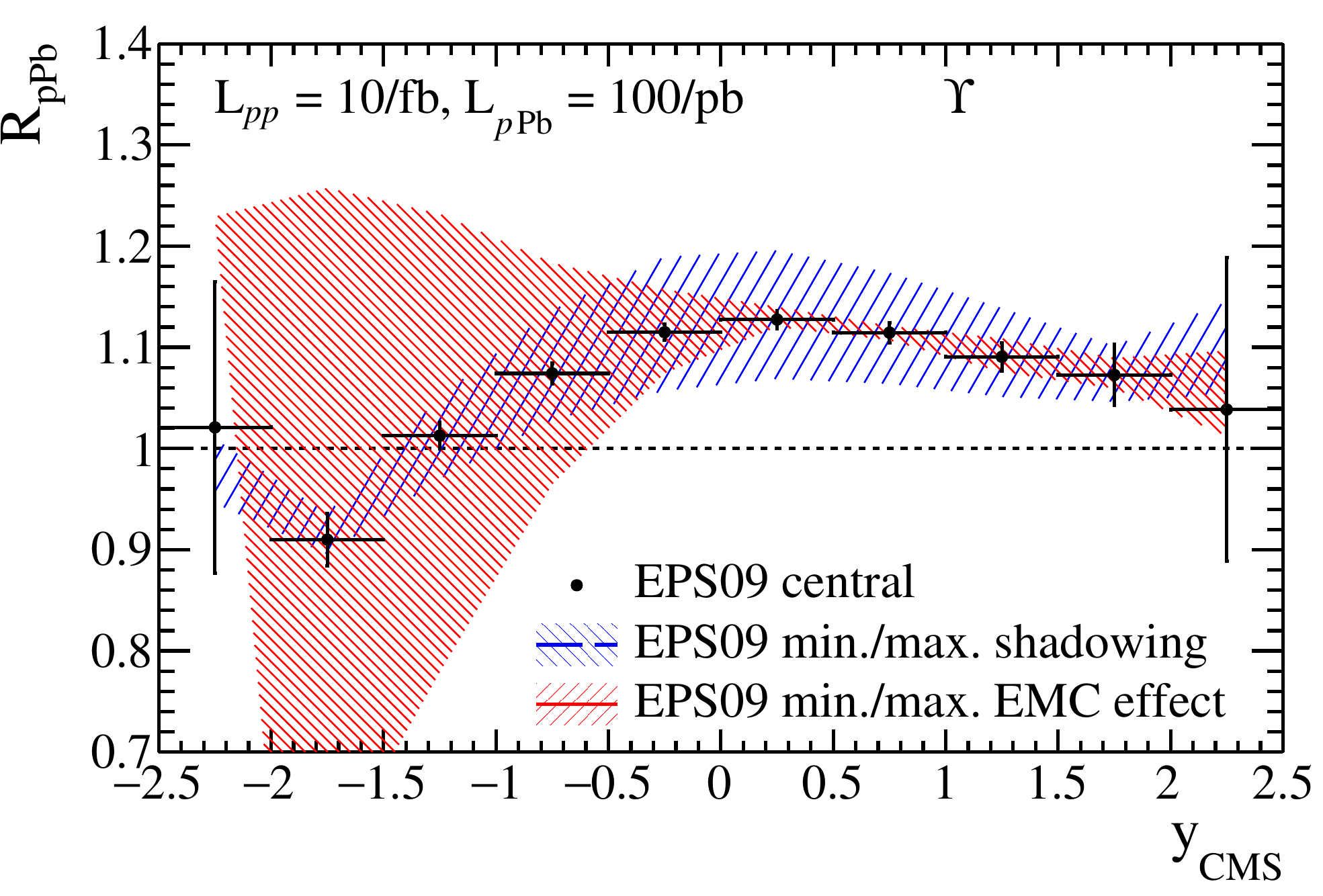}
		\includegraphics[width=0.9\columnwidth]{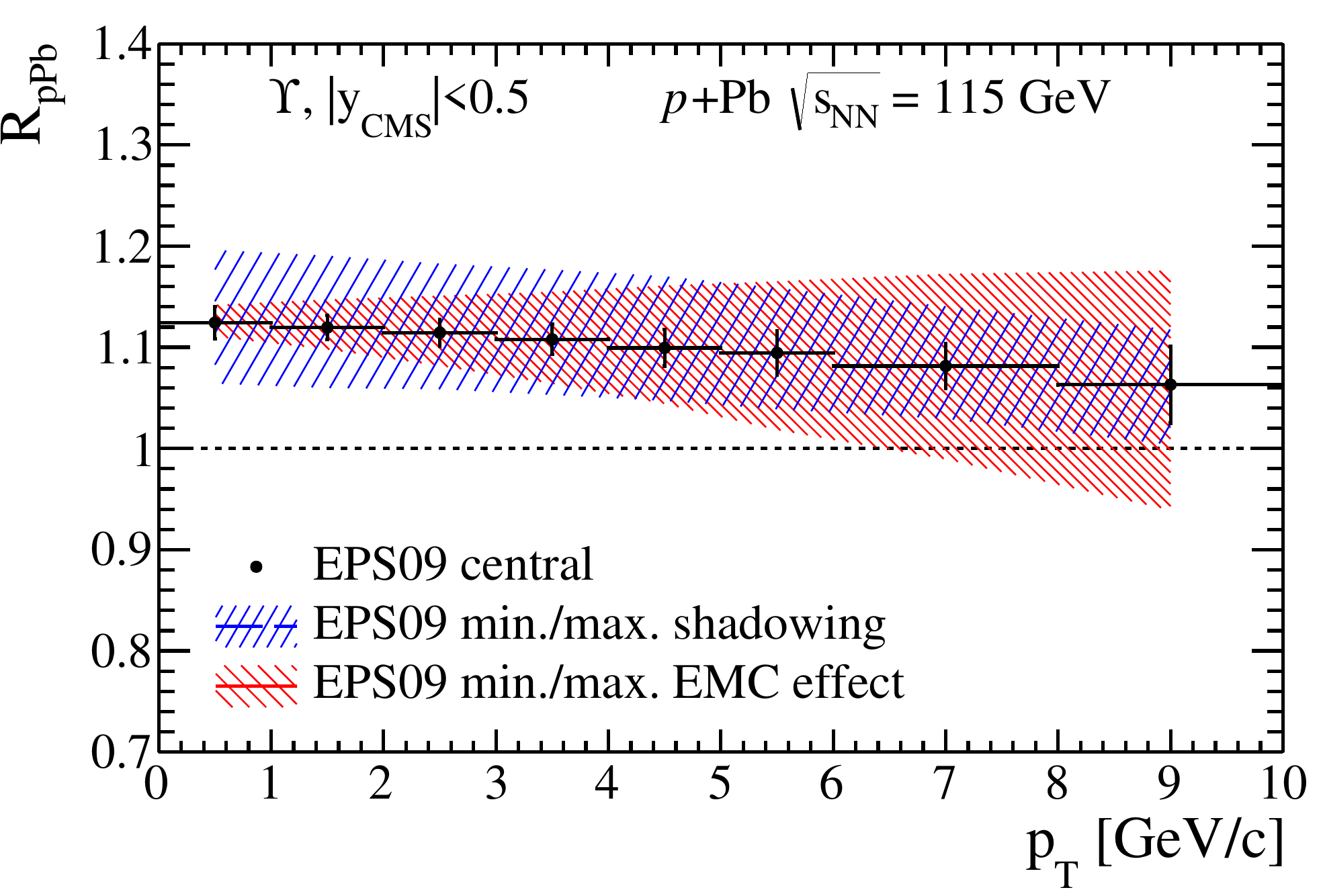}
		\caption{Nuclear modification factor for \ups  as a function of (a) $y_{\rm CMS}$ and (b) $p_\mathrm{T}$ in $p+\mathrm{Pb}$   collisions at $\snn=115$~GeV. The uncertainties attached to the central points are derived from the statistics to be collected with ${\cal L}_{p+p}=10$~fb$^{-1}$ and ${\cal L}_{p+\rm Pb}=100$~pb$^{-1}$.}
		\label{fig:rpa-upsi}
\end{figure*}

\begin{figure*}[ht]
		\centering
		\includegraphics[width=0.9\columnwidth]{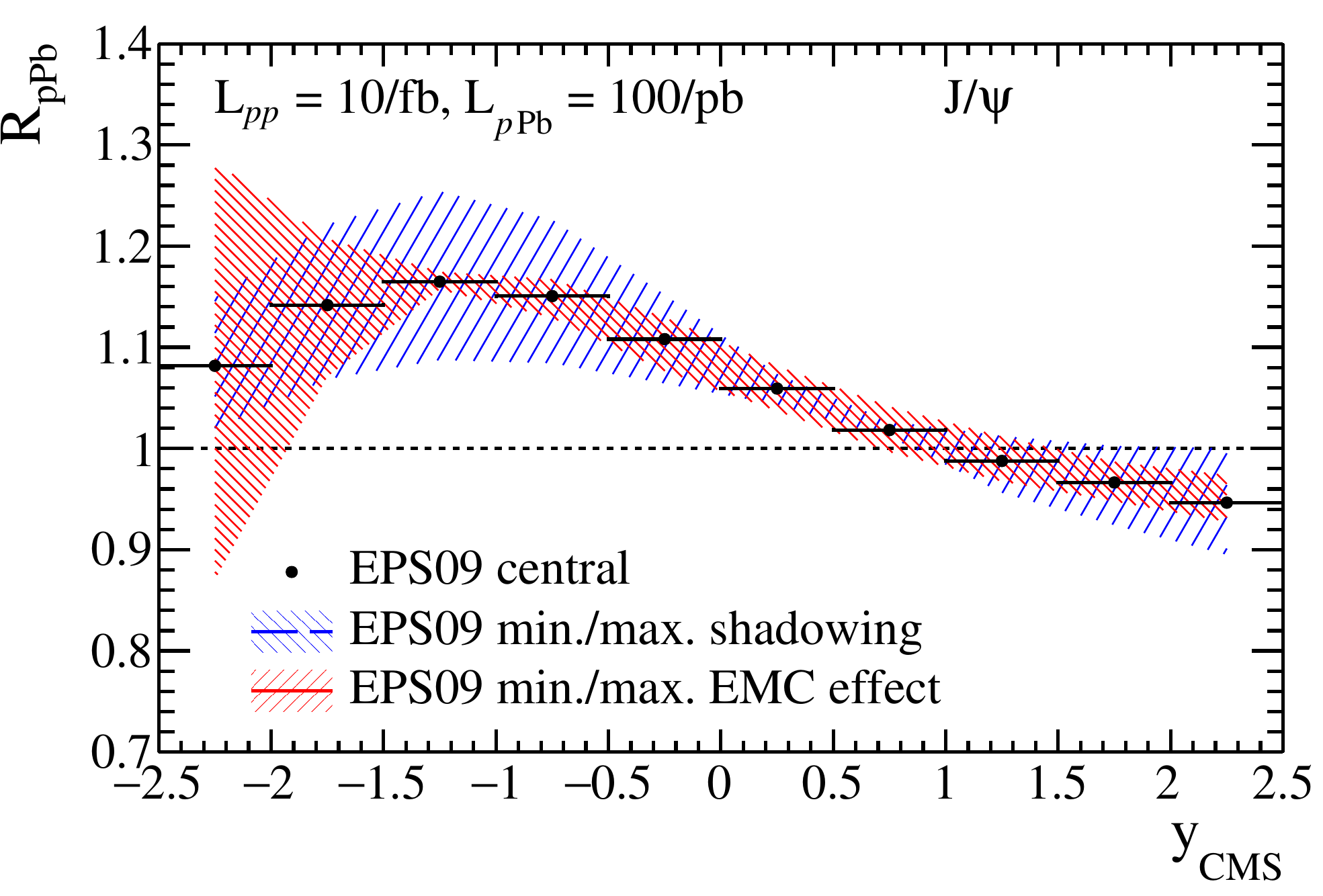}
		\includegraphics[width=0.9\columnwidth]{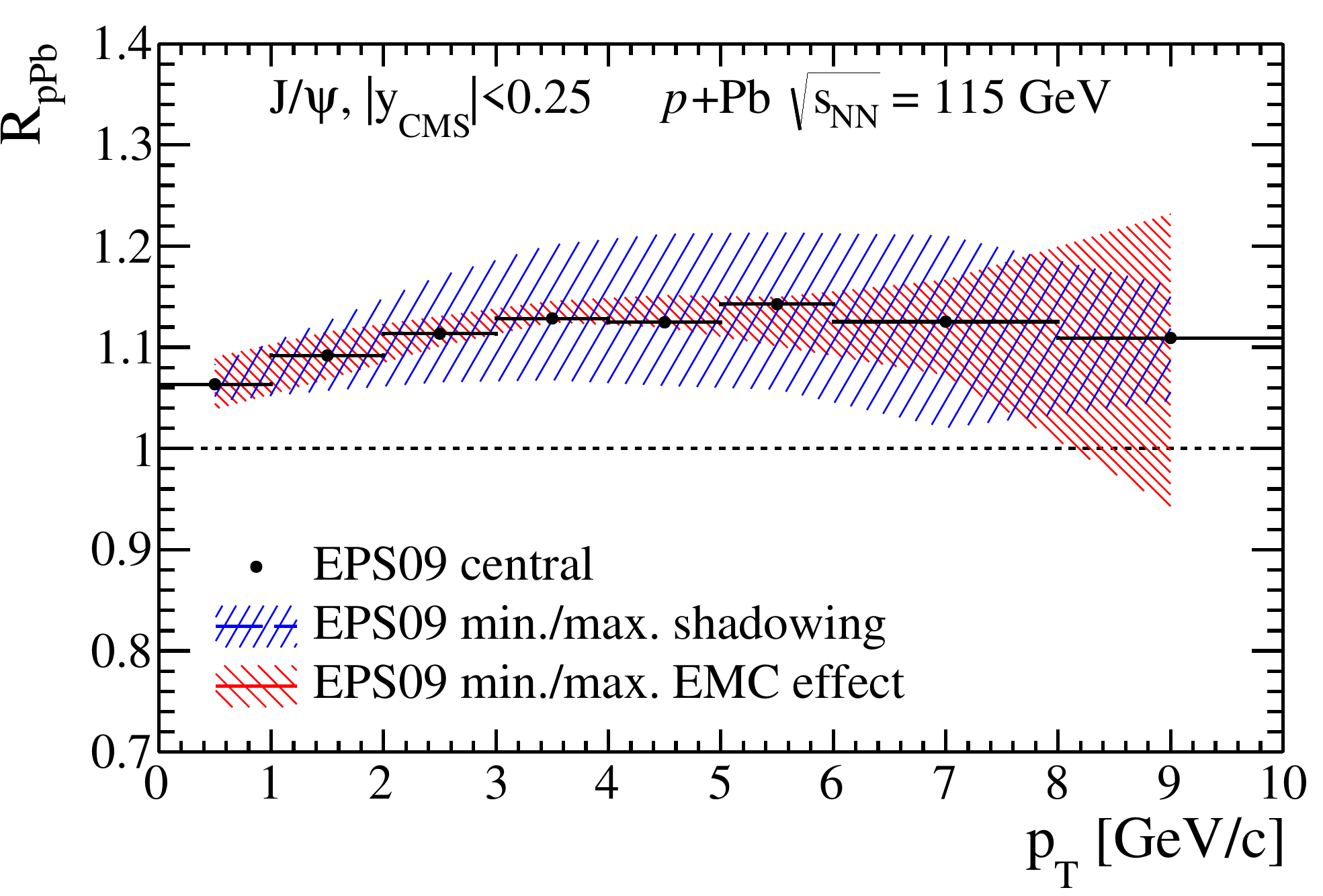}
		\caption{Nuclear modification factor for \jpsi as a function of (a) $y_{\rm CMS}$ and (b) $p_\mathrm{T}$ in $p+\mathrm{Pb}$ collisions at $\snn=115$~GeV. The uncertainties derived from the statistics to be collected with ${\cal L}_{p+p}=10$~fb$^{-1}$ and ${\cal L}_{p+\rm Pb}=100$~pb$^{-1}$ are smaller than the point size.}
		\label{fig:rpa-psi}
\end{figure*}

\begin{figure}[ht]
		\centering
		\includegraphics[width=0.9\columnwidth]{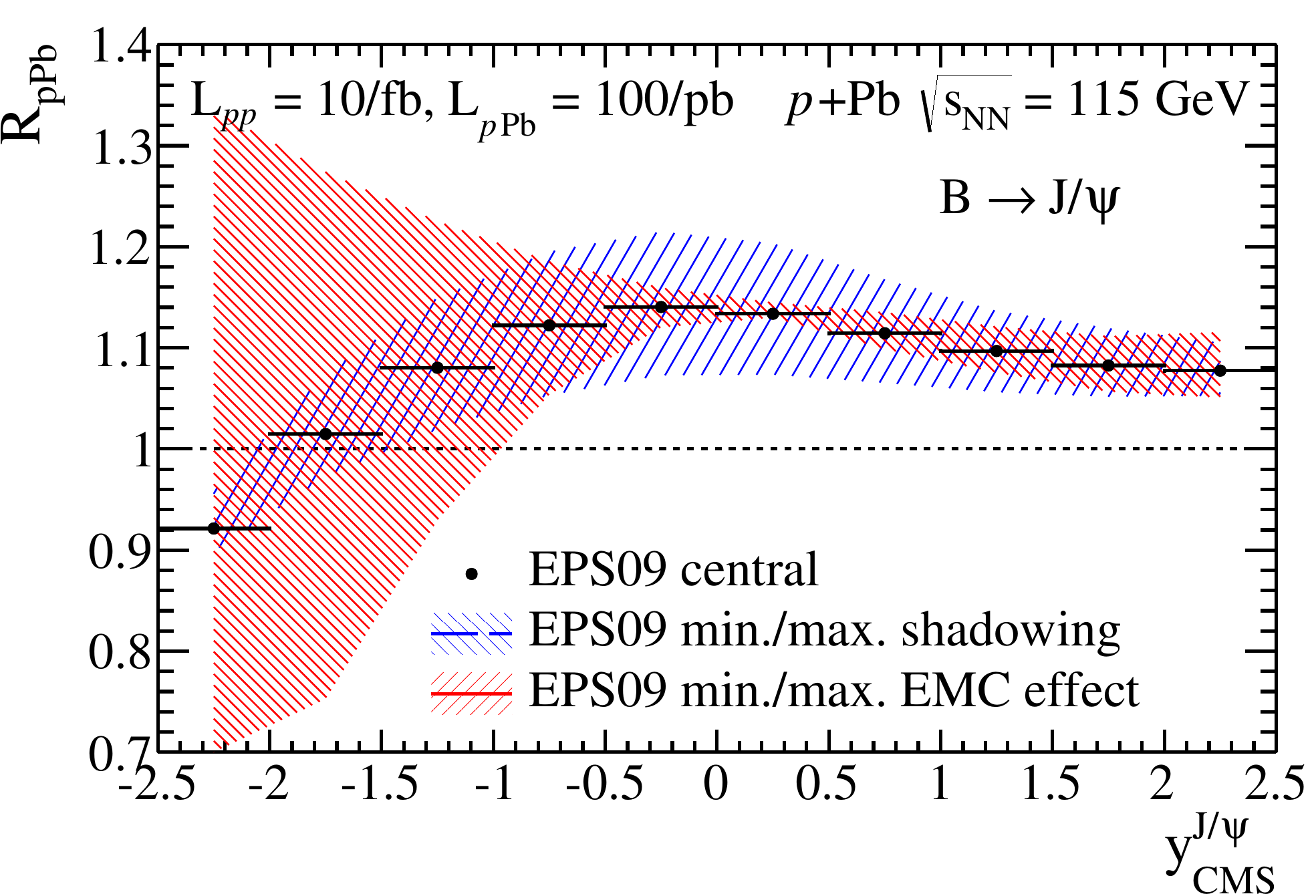}
		\caption{Nuclear modification factor for \jpsi from $b$  as a function of $y_{\rm CMS}$  in $p+\mathrm{Pb}$ collisions at $\snn=115$~GeV.
The uncertainties derived from the statistics to be collected with ${\cal L}_{p+p}=10$~fb$^{-1}$ and ${\cal L}_{p+\rm Pb}=100$~pb$^{-1}$ are smaller than the point size.}
		\label{fig:rpa-psifromb}
\end{figure}

One sees that looking at quarkonium production in $p+\mathrm{Pb}$  collisions at different backward rapidities 
allows one to look at quarkonia traversing the nuclear matter
at very different stages of their evolution. This effect could theoretically be studied by giving an ad-hoc 
rapidity dependence to the effective
absorption cross section, $\sigma^{\rm effective}_{\rm abs}$. This is left for a future study since, here,  we wish to consider only
the nPDF effects and the expected statistics. Other effects to be considered are the coherent energy loss~\cite{Arleo:2012hn} (expected to grow in the forward region)
and the rescattering by comovers~\cite{Ferreiro:2014bia} (expected to grow with the multiplicity along the J$/\psi$ direction).

Since we wish to assess the descriminating power of possible data to be taken with AFTER@LHC, we attribute
to the EPS09 central values statistical uncertainties which directly follow from the differential yields repectively expected
in $p+p$ and $p+\mathrm{Pb}$ collisions. For that we take an integrated luminosity of 10 fb$^{-1}$ for the
$p+p$ runs and 100 pb$^{-1}$ for the $p+\mathrm{Pb}$ runs, in accordance with the luminosities discussed above 
(see~\ct{tablumi}).
As this stage, we do not consider additional systematical uncertainties. This simplifying assumption 
could be lifted in a more detail study which would also take into account a possible detector acceptance 
(and related efficiencies) 
as done in the previous section. In particular, we do not expect that the rapidity region for $y_{\rm CMS} > 1.5$ would be easily accessible.

In \cf{fig:rpa-upsi}, we show the rapidity dependence of \RpPb\ for $\Upsilon$ and its $p_\mathrm{T}$ dependence near $y=0$.  
The million of $\Upsilon$ to be collected per year allows for the  measurement of a \RpPb\ with a much better precision than the gluon nPDF, nearly up
to $x \to 1$. In addition, one notes that the nuclear modification factor is certainly measurable up to $p_\mathrm{T} \simeq 10$ GeV/c. 

In \cf{fig:rpa-psi}, we also show the rapidity dependence of \RpPb\ for J$/\psi$ and its $p_\mathrm{T}$ dependence near $y=0$. In both cases, 
the luminosity to be taken in a year at AFTER@LHC yields to statistical uncertainties which are largely negligible 
as compared to the nPDF uncertainties -- the statistical uncertainties are not even visible on~\cf{fig:rpa-psi}. We except this to hold also for the $\psi(2S)$ although its yields are down by a factor of 100.

As aforementioned, the nPDFs do not account for all the expected nuclear matter effects. However, it is clear that combining the measurements of
$\Upsilon$, J$/\psi$ and $\psi(2S)$ for $-3< y_{\rm CMS} < 0$ (as a LHCb-like detector would do) will allow one to pin down the existence of
a possible gluon EMC and antishadowing effect. We also stress that the complications induced by a rapidity dependence 
of $\sigma^{\rm effective}_{\rm abs}$ could be avoided 
by the parallel measurement of \RpPb\ for non-prompt J$/\psi$ which can only be sensitive to the energy loss since the $b$ quark decay (weakly)
into the J$/\psi$, way outside the nucleus. \cf{fig:rpa-psifromb} shows that the trend is similar than for $\Upsilon$. 
Measuring the $p_\mathrm{T}$ dependence of \RpPb\ for prompt J$/\psi$ and $\Upsilon$ should also avoid the sensitivity on formation time effects.

\section{Prospects of P\lowercase{b}+A measurements at $\sqrt{s}=72$~GeV}
\label{sec:simuAA}

The charged particle multiplicity has been generated with 
the EPOS generator~\cite{Pierog:2013ria,Werner:2005jf} in different configurations: $\mathrm{Pb}+\mathrm{Pb}$ 
at $\snn=5.5$~TeV in collider mode, $\mathrm{Pb}+\mathrm{Ar}$, $\mathrm{Pb}+\mathrm{Xe}$ and 
$\mathrm{Pb}+\mathrm{Pb}$ at $\snn=72$~GeV in fixed-target mode. 
By comparing these three distributions 
in the pseudorapidity of the particle in the laboratory frame as shown in \cf{fig:mult-PbA}, 
one can conclude that the charged particle multiplicity 
in a fixed-target mode never exceeds the one obtained in $\mathrm{Pb}+\mathrm{Pb}$ collisions at $\snn=5.5$~TeV 
obtained in a collider mode in the full pseudorapidity range: a detector with the ALICE MFT+Muon detector~\cite{CERN-LHCC-2013-014} 
capability will be able to run in such conditions. 
Details studies are needed to evaluate up to which multiplicity a detector such as LHCb 
would be able to take good quality data.  

\begin{figure}[ht]
		\centering
		\includegraphics[width=\columnwidth]{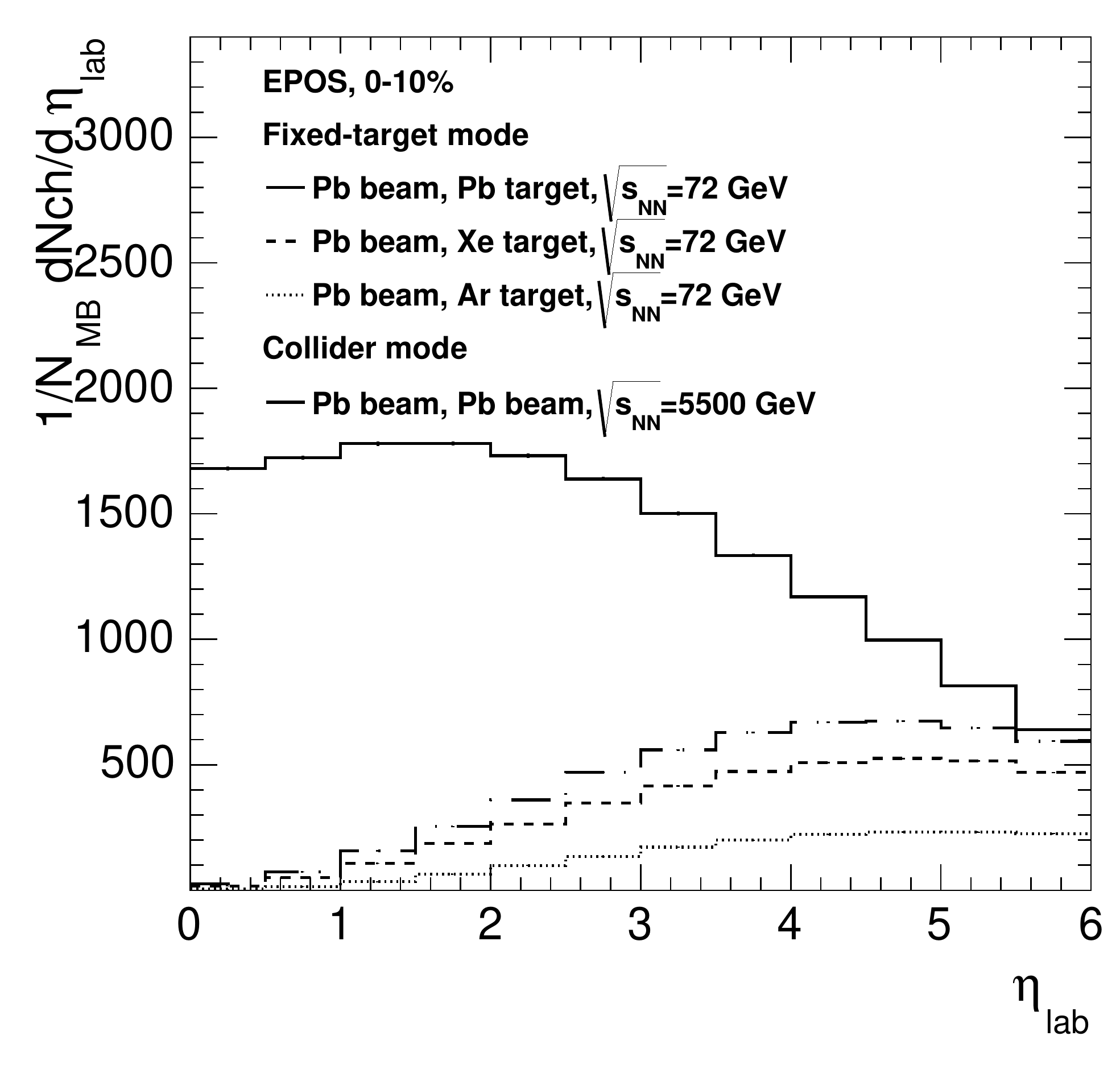}
		\caption{Averaged number of charged particles in  $A+A$  collisions as a function of the pseudorapidity in the laboratory frame.}
		\label{fig:mult-PbA}
\end{figure}

\section{Conclusion}
\label{sec:conclusions}

In summary, we have shown that in a fixed target mode with an integrated luminosity of 10 fb$^{-1}$, 
using 7 TeV LHC proton beam on a hydrogen target, and with a detector setup and performances similar 
to the LHCb detector, quarkonium studies in the dimuon decay channel can be performed over a wide 
transverse momentum range and rapidity in the center of mass from $\sim  -2.8$ for $J/\psi$ and 
$\psi(2S)$, and $\sim -2$ for $\Upsilon$ states, up to $\sim 0$. We have performed simulations 
of the dominant background sources contributing to the $\mu^{+}\mu^{-}$ invariant mass spectrum. 
The uncorrelated background was obtained using \Pythia\ generator and dimuons from correlated 
background sources: $c\bar{c}$, $b\bar{b}$ and Drell-Yan, were simulated using both \HELACOnia\ 
and \Pythia\ generators. The estimated background level allows for $J/\psi$,
 $\psi(2S)$, $\Upsilon(1S)$, $\Upsilon(2S)$ and $\Upsilon(3S)$ measurements in the
dimuon decay channel with good signal to background ratios. 

These simulations set the stage for further ones including, on the one hand, the 
detection of photon from $P$ wave or $\eta_c$ decay or from the production of a $J/\psi+\gamma$ pair,
 whose studies at low transverse momentum can provide
important insight on the gluon transverse dynamics~\cite{Boer:2012bt,Ma:2012hh,Dunnen:2014eta,Lansberg:2015hla} 
and, on the other hand, the large combinatorial background typical of 
$p+A$ and $A+A$ collisions in which the study of excited quarkonium at AFTER@LHC energies is of 
paramount importance~\cite{Brodsky:2012vg,Lansberg:2012kf}. 
We note that the {\sc Delphes}~\cite{deFavereau:2013fsa} framework seems particularly 
well suited to account for the photon detectability in such prospective studies.

Along our investigations, we have also noted that main source of dimuons around the $\Upsilon(nS)$ masses is from 
the Drell-Yan process (see~\cf{fig:QuarkoniaWithBkg} (right)). This gives us great confidence that the corresponding
cross section can easily be extracted in this mass region in $p+p$ collisions, a fortiori 
with a vertex detector allowing for tagging the heavy-flavour muons. 
We therefore consider that the single-spin asymetries for Drell-Yan pair 
production can indeed be extracted using a light polarised target. Motivations for such
studies are discussed in \cite{Liu:2012vn,Kanazawa:2015fia,Anselmino:2015eoa}.
Quarkonium polarisation measurement are of course also possible given the large statistical samples.

As regards the case of $p+A$ collisions, we have had a first look at the charged particle
multiplicities as a function of the laboratory pseudo-rapidity. We have found out that, for all
the possible fixed target modes, $p+$Pb, Pb+H, Pb+Pb, these are smaller than the ones
reached in the collider modes where the LHCb was used ($p+$Pb and Pb+$p$ at 5 TeV). We therefore believe that a detector
with similar characteristics as compared to LHCb can very well be used in the fixed-target 
mode~\footnote{Our observation is obviously supported by the preliminary analysis of the LHCb-SMOG data taken during the
 pilot run of $p^{+}$ beam (Pb beam) on a Neon gas target from 2012 (2013) at a c.m.s energy of $\sqrt{s_{NN}}$ = 87 GeV (54 GeV)~\cite{LHCb:2012aka}.}.

In view of the above, we have evaluated the impact --and its uncertainty-- on 
the nuclear modification of the gluon densities  on prompt and non-prompt $J/\psi$ and $\Upsilon$ in form of \RpPb. 
We have found that the measurements at backward rapidities allows one to search for the 
gluon antishadowing, the gluon EMC effect and even the Fermi motion effect on the gluons with
unheard of statistical precisions. 
The statistics are large enough to perform such measurement with the $\psi(2S)$ and probably also with
$\Upsilon(2S)$ and $\Upsilon(3S)$ allowing for thorough investigations of formation time effect of the meson
propagating in the nuclear matter. Overall, our results confirm the great potential of AFTER@LHC for heavy-quark and
quarkonium physics.

\section*{Acknowledgements}

We thank C.~Da Silva, D.~d'Enterria,  E.G.~Ferreiro, R.~Mikkelsen, S.~Porteboeuf-Houssais, A.~Rakotozafindrabe, P.~Robbe, M.~Selvaggi, 
M.~Schmelling, P.~Skands and Z.~Yang for important and stimulating discussions. 
This research was supported in part by the ERC grant 291377 "LHCtheory: Theoretical predictions and analyses of LHC physics: 
advancing the precision frontier", by the COPIN-IN2P3 Agreement, by the French P2I0 Excellence Laboratory, by the French CNRS via the grants PICS-06149 Torino-IPNO,
FCPPL-Quarkonium4AFTER \& PEPS4AFTER2, by the European social fund within the framework of realizing the project ``Support of 
inter-sectoral mobility and quality enhancement of research teams at Czech Technical University in Prague'', CZ.1.07/2.3.00/30.0034, by Grant Agency of the Czech Republic, grant No.13-20841S and by the Foundation for Polish Science Grant HOMING PLUS/2013-7/8.

\bibliographystyle{utphys}

\bibliography{biblio-170415}


\end{document}